\documentclass[
    twocolumn,
    notitlepage,
    pra,
    showpacs,
    superscriptaddress,
    amssymb,
    amsmath,
    amsmath,
    aps,
    10pt,]{revtex4-2}
\usepackage{dsfont}
\usepackage{lipsum}
\usepackage{graphicx}
\usepackage{epstopdf}
\usepackage{dcolumn}
\usepackage{bm}
\usepackage{physics}
\usepackage{xcolor}
\usepackage{amssymb}
\usepackage{gensymb}
\usepackage{amsmath}
\usepackage{soul}
\usepackage[colorlinks=true, linkcolor=blue, urlcolor=blue, citecolor=blue]{hyperref}
\usepackage{siunitx}
\usepackage{makecell}
\usepackage{tabularx}
\usepackage{booktabs}
\usepackage{mathtools}

\usepackage{xcolor}
\usepackage{ marvosym }
\usepackage{pifont}
\newcommand{\cmark}{\ding{51}}%
\newcommand{\xmark}{\ding{55}}%
\usepackage{tikz, ifthen}

\usetikzlibrary{arrows,shapes,shadows,fit,positioning, calc, backgrounds,decorations.pathmorphing}
\tikzset{
diagonal fill/.style n args={5}{fill=#1, draw, label=center:{#5}, minimum size=1.00cm, path picture={
\fill[#2, sharp corners] (path picture bounding box.south west) -| 
  (path picture bounding box.south east) -- (path picture bounding box.center) -- cycle;
\fill[#3, sharp corners] (path picture bounding box.north west) -| 
  (path picture bounding box.south west) -- (path picture bounding box.center) -- cycle;
\fill[#4, sharp corners] (path picture bounding box.north east) -| 
  (path picture bounding box.north west) -- (path picture bounding box.center) -- cycle;
}}
}

\usepackage{acro}
\DeclareAcronym{mpo}{ short = MPO, long = matrix product operator, }
\DeclareAcronym{peps}{ short = PEPS, long = projected entangled pair state, }
\DeclareAcronym{ipeps}{ short = iPEPS, long = infinite projected entangled state operator, }
\DeclareAcronym{pepo}{ short = PEPO, long = projected entangled pair operator, }
\DeclareAcronym{ipepo}{ short = iPEPO, long = infinite projected entangled pair operator, }
\DeclareAcronym{fsa}{ short = FSA, long = finite signaling agent, }
\DeclareAcronym{fsm}{ short = FSM, long = finite state machine, }
\DeclareAcronym{itrsu}{ short = itrSU, long = iterative simple update, }
\DeclareAcronym{vumps}{ short = VUMPS, long = variational uniform matrix product state,}
\DeclareAcronym{cptp}{ short = CPTP, long = completely positive trace preserving, }
\DeclareAcronym{mps}{ short = MPS, long = matrix product state, }
\DeclareAcronym{cpepo}{ short = tePEPO, long = time-evolving PEPO, }

\usepackage{newfloat}
\usepackage{algpseudocode}
\DeclareFloatingEnvironment[
    fileext=loa,
    listname=List of Algorithms,
    name=ALGORITHM,
    placement=tbhp,
]{algorithm}

\usepackage[T1]{fontenc} 

\newcommand*{\la}{\langle}
\newcommand*{\ra}{\rangle}
\newcommand*{\bb}[1]{\left(  #1 \right)}

\newcommand{\kmax}{k_{\max}}
\newcommand{\nmax}{n_{\max}}
\newcommand{\tol}{\mathrm{tol}}

\newcommand{\vecb}[1]{| #1 \ra \! \ra}
\newcommand{\vecp}{\vecb{\rho}}

\newcommand{\sx}[1]{\hat{\sigma}^{x}_{#1}}

\newcommand{\sz}[1]{\hat{\sigma}^{z}_{#1}}

\newcommand{\I}{\mathds{1}}

\newcommand{\Wk}{W^{[k]}(\tau)}

\newcommand{\WI}{$\mathrm{W}^{\mathrm{I}}$ }
\newcommand{\WII}{$\mathrm{W}^{\mathrm{II}}$ }

\DeclareMathOperator{\vecc}{vec}

\newcommand{\fsmboxx}[4]{
\tikzset{
diagonal fill/.style n args={5}{fill=##1, draw, label=center:{}, minimum size=0.5cm, path picture={
\fill[##2, sharp corners] (path picture bounding box.south west) -| 
  (path picture bounding box.south east) -- (path picture bounding box.center) -- cycle;
\fill[##3, sharp corners] (path picture bounding box.north west) -| 
  (path picture bounding box.south west) -- (path picture bounding box.center) -- cycle;
\fill[##4, sharp corners] (path picture bounding box.north east) -| 
  (path picture bounding box.north west) -- (path picture bounding box.center) -- cycle;
}}
}
\pgfmathsetmacro{\cole}{ ifthenelse(#1 == 0, "white", #1 == 1? "red!30" : "blue!30")}
\pgfmathsetmacro{\cols}{ ifthenelse(#2 == 0, "white", #2 == 1? "red!30" : "blue!30")}
\pgfmathsetmacro{\colw}{ ifthenelse(#3 == 0, "white", #3 == 1? "red!30" : "blue!30")}
\pgfmathsetmacro{\coln}{ ifthenelse(#4 == 0, "white", #4 == 1? "red!30" : "blue!30")}
\node[diagonal fill={\cole}{\cols}{\colw}{\coln}{}] at (0,0) {};
}

\newcommand{\fsmbox}[4]{\vcenter{\hbox{
\begin{tikzpicture}
  \fsmboxx{#1}{#2}{#3}{#4}
\end{tikzpicture}
}}}

\newcommand{\fsmdot}[4]{\vcenter{\hbox{
\begin{tikzpicture}
  \fsmboxx{#1}{#2}{#3}{#4}
  \node[fill, black, circle, inner sep = 1pt] at (0,0){};
\end{tikzpicture}
}}}

\newcommand{\isep}{\, \ldotp \ldotp \,}

\newcommand{\secref}[1]{Sec.~\ref{#1}}
\newcommand{\figref}[1]{Fig.~\ref{#1}}
\newcommand{\tabref}[1]{Table~\ref{#1}}

\newcommand{\affCFT}{Center for Theoretical Physics, Polish Academy of Sciences, Al. Lotników 32/46, 02-668 Warsaw, Poland}
\newcommand{\affLCN}{London Centre of Nanotechnology, 9 Gordon St, London WC1H 0AH, United Kingdom}
\newcommand{\affUCL}{Department of Physics and Astronomy, University College London, Gower Street, London, WC1E 6BT, United Kingdom}

\preprint{APS/123-QED}

\begin{document}

\title{Efficient Time Evolution of 2D Open-Quantum Lattice Models with Long-Range Interactions using Tensor Networks}

\author{J. Dunham}
\altaffiliation{jdunham@flatironinstitute.org}
\affiliation{\affUCL}
\affiliation{\affLCN}
\affiliation{\affCFT}
\author{M. H. Szyma\'nska}
\affiliation{\affUCL}

\date{November 20, 2025}

\begin{abstract}
    Simulating many-body open quantum systems is an extremely challenging problem, with methods often restricted to either models with nearest-neighbor interactions or semi-classical approximations.
    In particular, modeling two-dimensional systems with realistic long-range interactions, in addition to dissipation, is of vital importance to the development of modern  quantum computing and simulation platforms. 
	In this paper, we present a construction of the time-evolution operator, as a projected entangled pair operator (denoted tePEPO), that can be used to evolve a tensor network ansatz through time. 
    Interactions beyond nearest-neighbor, including interactions between sites not collinear in the lattice, can be represented efficiently as a tePEPO.
	Furthermore, we obtain approximations to realistic radial long-range interactions decaying with a power-law, that give accurate results with small tePEPO bond dimension.
    Finally, we consider a physical example of a Rydberg atom Hamiltonian with long-range dipolar interactions, and show evidence of a dipole-dipole blockading effect in presence of dissipation.
     This work demonstrates the applicability of tensor networks to two-dimensional systems widely studied in experiments, but previously inaccessible to non-semi-classical methods.
\end{abstract}

\maketitle


\section{Introduction}

Precise control of arrays of interacting quantum objects is a task central to the development of modern quantum technologies such as quantum computers and  simulators.
Ideally, these \emph{many-body} quantum systems would be perfectly isolated from their environment, operating as  closed systems; however in the real world this is practically impossible to achieve. While strides have been made in mitigating decoherence caused by external effects~\cite{schlosshauer_quantum_2019} in such \emph{open} quantum systems, many novel phenomena can actually emerge from the interplay between a systems coherent evolution, external drive, and dissipative effects, with no counterpart found in closed systems.
%
In addition, many-body open quantum systems have been proposed as a computation resource~\cite{verstraete_quantum_2009}, for quantum state preparation~\cite{cole_dissipative_2021, yang_dissipative_2023, wampler_absorbing_2024, zhan_rapid_2025}, to aid quantum error correction~\cite{gertler_protecting_2021}, and as store of energy in the form of a quantum battery~\cite{carrasco_collective_2022}, among others. Furthermore, the rapid advancement in engineering affords us unparalleled control of such driven-dissipative effects in platforms such as exciton-polaritons~\cite{amo_exciton-polaritons_2016}, ultracold atoms~\cite{schafer_tools_2020}, and circuit- and cavity-QED~\cite{blais_circuit_2021, walther_cavity_2006}, to name a few.
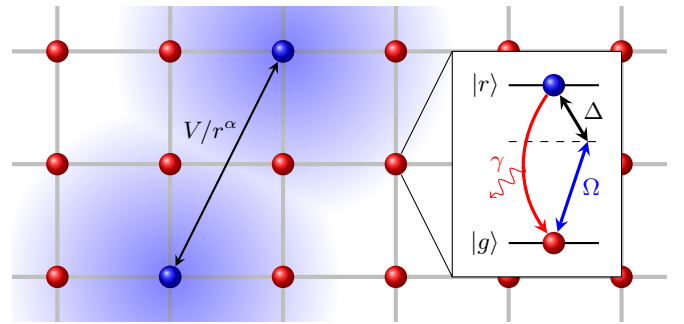
\begin{figure}[b]
\centering
\begin{tikzpicture}[decoration=snake, xscale=1.5, yscale=1.5]
\begin{scope}[blend group=darken]
     \clip (0.6,0.6) rectangle (6.4,3.4);
     \foreach \x in {1,2,3}{
        \draw[gray!50!white, ultra thick] (0.5, \x) to (6.5, \x);
     }
     \foreach \x in {1,2,3,4,5,6}{
        \draw[gray!50!white, ultra thick] (\x, 0.5) to (\x, 4);
     }
     \shade[blue, inner color=blue!50, outer color=blue!0] (3,3) circle (1.5);
     \shade[blue, inner color=blue!50, outer color=blue!0] (2,1) circle (1.5);
\end{scope}
\foreach \x [count=\xi] in {1,2, 3, 4, 5, 6}{
  \foreach \y [count=\yi] in {1,2,3}{
    \node[ball color=red, circle, inner sep=3pt] (\xi\yi) at (\x,\y) {};
  }
}
\node[ball color=blue, circle, inner sep=3pt] (1) at (33) {};
\node[ball color=blue, circle, inner sep=3pt] (2) at (21) {};
\draw[thick, stealth-stealth] (1) to node[midway, yshift=12pt, xshift=-6pt] {$V/r^{\alpha}$}(2) ;
\fill[white, draw=black] (4.5,1.0) rectangle (6,3);
\draw[thick] (5,2.7) node[left] {$\ket{r}$} to (5.8, 2.7); 
\draw[thick] (5,1.3) node[left] {$\ket{g}$} to (5.8, 1.3);
\draw[dashed] (5, 2.2) to (5.8, 2.2);
\node[ball color=red, circle, inner sep=3pt] (g) at (5.4, 1.3) {};
\node[ball color=blue, circle, inner sep=3pt] (e) at (5.4,2.7) {};
\draw (42) to (4.5,1.0);
\draw (42) to (4.5,3);
\coordinate (m) at (5.7, 2.2);
\draw[very thick, blue, stealth-stealth] (g) to [bend left=0] node[midway, right] {$\Omega$} (m);
\draw[very thick, stealth-stealth] (m) to node[midway, right, yshift=3pt] {$\Delta$} (e);
\draw[very thick, red, stealth-] (g) to  [bend left=35] node[midway, label=left:{$\gamma$}] (gamma) {} (e) ;
\draw[decorate, red, segment length=2.1mm, ->] (gamma.center) -- ++ (-0.3,-0.3);
\end{tikzpicture}
\caption{{\bf Example of a many-body open quantum system.} Atoms in  excited Rydberg states $\ket{r}$ with strong, long-range repulsive interactions decaying as a power law $r^{-\alpha}$, typically of van der Waals $\alpha = 6$ or dipolar $\alpha = 3$ form. 
Atoms are promoted from the ground state $
\ket{g}$ to a highly-excited Rydberg state $\ket{r}$ via a laser drive with frequency $\Omega$ and detuning $\Delta$.
Atoms can then decay $\ket{r}\to \ket{g}$ incoherently at a rate $\gamma$.}
\end{figure}

In light of this, the task of modeling many-body open quantum systems is vital to the development of quantum technologies in both the near- and long-term, while still remaining one of the most formidable challenges in computational physics. 
Many-body quantum systems occupy a Hilbert space that is exponentially large with respect to the number of particles (for example), precluding an exact simulation of more then only a few.
Furthermore, the state of an open quantum system is typically expressed as a \emph{density matrix}, leading to an additional quadratic penalty compared to the state vector description of a closed system.
In some cases, exact solutions have been found~\cite{foss-feig_solvable_2017,mcdonald_exact_2022}, but these are limited, and while analytical techniques such as the Keldysh formalism~\cite{sieberer_keldysh_2016} can provide insights beyond the mean-field approximation~\cite{maghrebi_nonequilibrium_2016}, for the most part, sophisticated numerical techniques must be employed. 
Moreover, the wealth of numerical methods in the literature of closed systems are not always transferable, necessitating the development of techniques specific to open quantum systems.

To this end, a variety of numerical approaches have emerged. Variational Monte Carlo methods~\cite{nagy_driven-dissipative_2018} have been used to find steady-states and dynamics of Lindblad equations using a neural network~\cite{hartmann_neural-network_2019, nagy_variational_2019,vicentini_variational_2019,yoshioka_constructing_2019, kothe_liouville-space_2024}, and more recently, a tensor network ansatz~\cite{hryniuk_tensor-network-based_2024, hryniuk_variational_2025}.
Unfortunately, as discussed in Ref.~\cite{nagy_driven-dissipative_2018}, the complex-valued eigenvalues of the associated Liouvillian super-operator can, in some cases, lead to the infamous sign problem~\cite{troyer_computational_2005} in such methods.
Methods using the phase space representation of the density matrix, such as the truncated Wigner approximation~\cite{mink_hybrid_2022,singh_driven-dissipative_2022,huber_realistic_2022} and Positive-P are highly scalable, however the former is semi-classical, failing to capture all physical aspects of bosonic systems~\cite{vanregemortel_spontaneous_2017}, and the latter, while particularly suited to system with dissipation~\cite{deuar_fully_2021}, has yet to be formulated for spins.

For closed systems, \emph{tensor network methods}, such as the celebrated density matrix renormalization group (DMRG) algorithm~\cite{white_density-matrix_1993, schollwock_density-matrix_2011}, have been enormously successful in the treatment of strongly-correlated many-body quantum lattice models. 
Such methods represent the quantum state as a network of tensors, which geometry is intrinsically related to entanglement structure of the underlying state. 
Tensor networks then parameterize an exponentially smaller, physically relevant corner of the total Hilbert space. 
The dimension of the indices that connect these tensors in the network, known as the bond dimension, acts a parameter that controls the expressibility of the ansatz. 
For open quantum systems, methods based on tensor networks are attractive as they tend to not impose additional approximations on the equations of motion, in contrast to semi-classical approaches, and allow direct access to the density matrix.
This enables  one to compute entanglement measures, such as mutual information negativity, not necessarily available to phase-space methods.
Moreover, increasing the bond dimension allows one to, in principle, systematically improve the simulation by allowing the ansatz to capture more of the quantum correlations and entanglement present in the system.

For the most part, the majority of tensor network works in this field have been limited to one-dimensional systems~\cite{verstraete_matrix_2004, cui_variational_2015, mascarenhas_matrix-product-operator_2015, werner_positive_2016, gangat_steady_2017, jaschke_one-dimensional_2018, sulz_numerical_2024, godinez-ramirez_riemannian_2024, saiphet_simulation_2025, allen_simulating_2025}.
For two-dimensional open quantum systems, studies are scarce, and limited to nearest-neighbor interactions~\cite{kshetrimayum_simple_2017, keever_stable_2021, kilda_stability_2021}.
They employ the two-dimensional time-evolving block decimation (TEBD)~\cite{vidal_efficient_2004,verstraete_matrix_2004, white_real-time_2004, daley_time-dependent_2004} algorithm to solve the Lindblad master equation, a method most applicable to models with only nearest-neighbor interactions.
Studies moving beyond the nearest-neighbor case are rarer still; the extent that tensor networks can be applied to Lindblad time evolution beyond the nearest-neighbor case remains largely unexplored.
Some progress has been made to this end, albeit restricted to one-dimensional systems: long-range interactions have been modeled on finite chain using tree-tensor networks (TTN)~\cite{sulz_numerical_2024, saiphet_simulation_2025}, however TTN are not able to capture the two-dimensional area-law, severely limiting the expressibility of such an ansatz.
Recently, Monte Carlo with a matrix-product state ansatz  capable of studying long-range interactions was proposed~\cite{hryniuk_tensor-network-based_2024, hryniuk_variational_2025}, however this requires contracting finite-sized two-dimensional tensor networks, a task in the $\sharp\mathrm{P}$ complexity class~\cite{haferkamp_contracting_2020}.

Evidently, the study of open quantum systems with long-range models is an under-explored area of research, despite their relevance to current experimental platforms in quantum computing and simulation~\cite{raghunandan_high-density_2018,bernien_probing_2017,ebadi_quantum_2021}.
Furthermore, it was shown using a variational approach that a full treatment of the long-range tails of dissipative model gives significant corrections compared to considering an effective short-range model~\cite{kazemi_driven-dissipative_2023}.
In light of this, the contribution of this paper is the following: we present a recipe for constructing tensor network representations of time-evolution operators for two-dimensional lattice models with interactions beyond nearest-neighbor.
This represents a generalization from one dimension to higher dimensions of what are known as the \WI and \WII methods~\cite{zaletel_time-evolving_2015}, which are described in more detail later, and a partial generalization of the associated higher-order cluster expansions~\cite{vandamme_efficient_2024}.
The resulting tensor network time-evolution operators take the form of a \ac{pepo}, and are denoted \ac{cpepo}.
The constructed operators can then be used to evolve an \ac{ipepo} representation of the density matrix through time via a series of apply-and-truncate steps.

We find that realistic 2D power-law interactions can be achieved by considering sums of Gaussian terms, each representable by \ac{cpepo}, in analogy to approximations of 1D power-law with sums of exponential~\cite{pirvu_matrix_2010}. 
Moreover, we show that good agreement with exact solutions can be achieved with only a modest number of terms in this summation.
This then enables the simulation of long-range models, which short-range equivalents prove difficult to simulate with existing tensor network method under equivalent approximations, thus demonstrating an increase in the applicability of such methods to problems in many-body quantum physics.
The primary motivation and subsequent application of the methods in this paper is to the Markovian evolution of an open quantum systems, however the construction of the \ac{cpepo} is generic and also applicable to real and imaginary time evolution of closed systems.  


The paper is structured as follows.
We first provide some additional technical context to this work in \secref{sec:bg}.
In \secref{sec:op}, we introduce the recipe used to construct two-dimensional tensor-network operator representations of the size-extensive approximation to the time-evolution~\cite{zaletel_time-evolving_2015}, henceforth denoted the \ac{cpepo}.
In \secref{sec:lr}, we then describe how to approximate interactions decaying with a power-law as a sum of Gaussian functions, which have \ac{cpepo} representations.
\secref{sec:app} presents an efficient algorithm that solves for the time dynamics of the Lindblad equation by repeatedly applying a \ac{cpepo} to a \ac{ipepo} density matrix ansatz. 
To benchmark, this algorithm is then applied to an exactly solvable dissipative Ising model in  \secref{subsec:exact}, where it is shown to give accurate results beyond mean field in the strongly dissipative regime.  
We also show convergence beyond the exactly solvable regime of this model in~\secref{subsec:beyond}.
Finally, a dissipative Rydberg Hamiltonian with dipolar interactions is considered in \secref{subsec:rydberg}, where we show the characteristic dipole-dipole blockade effect.
We conclude with a discussion, and suggestions of future research directions.

\section{\label{sec:bg}Background}

This paper considers the problem of developing tensor network methods capable of simulating the time-evolution of long-range interacting systems. 
While readily applicable to real- and imaginary-time evolution of closed systems, we are interested in open quantum system obeying the following Markovian master equation ($\hbar = 1$):
\begin{equation}\label{eq:lindblad}
	\frac{d\rho}{dt} = \mathcal{L}(\rho) =- i [H, \rho] + \mathcal{D}(\rho)
	,\end{equation}
known as the Gorini–Kossakowski–Sudarshan–Lindblad (GKSL) equation~\cite{breuer_theory_2007}.
The Hamiltonian $H$ of the system governs the coherent dynamics of the quantum state $\rho$, and the dissipator $\mathcal{D}$, which models the system-environment coupling,
\begin{equation}
	\mathcal{D}(\rho) = \sum_{k} L_k \rho L_k^{\dagger} -
	\frac{1}{2}\bb{L^{\dagger}_k L_k \rho + \rho L^{\dagger}_k L_k}
,\end{equation}
is defined in terms of a set of often-phenomenological jump operators $\{L_k\}$.
The solution to equation~\eqref{eq:lindblad} is then the family of completely-positive trace-preserving (CPTP) maps obtained via exponentiation of the Liouvillian super-operator $\mathcal{L}$, i.e. 
\begin{equation}\label{eq:dyn-map}
    e^{t\mathcal{L}}: \rho(0) \mapsto \rho(t).
\end{equation}
Specifically, we seek a tensor network operator representation of~\eqref{eq:dyn-map} for two-dimensional lattice systems with interactions \emph{beyond} the simple nearest-neighbor case.
It is therefore useful to provide a summary of the major technical progress made to this end, and highlight some limitations of existing approaches.

In one-dimension, beyond the context of open quantum systems, interactions decaying with the exponential of distance admit efficient representations as \ac{mpo}, which can then be used as a basis to approximate other decay functions~\cite{crosswhite_finite_2008, pirvu_matrix_2010}.
This idea was applied to time dynamics in a size-extensive way by employing techniques from finite-state automata~\cite{crosswhite_finite_2008, zaletel_time-evolving_2015}.
By extending the cluster expansion approach~\cite{vanhecke_symmetric_2021}, this idea was generalized further:~\cite{vandamme_efficient_2024} presents an algorithm that generates an extensive time-evolution matrix product operator of arbitrary order and, through a series of extension and compression steps, can be made optimal in terms of the operator bond-dimension.
Provided the operator can be written as an \ac{mpo} in so-called \emph{regular form}~\cite{parker_local_2020}, the algorithms in \cite{vandamme_efficient_2024} can be applied.
The generalization to Lindblad dynamics is straightforward via vectorization~\cite{[See ][ for an example of how to construct the required MPO form of a Liouvillian.]jaschke_one-dimensional_2018} and is of particular interest as completely-positive and trace-preserving (CPTP) Trotter decompositions have not been found beyond second order~\cite{werner_positive_2016}.

Such \ac{mpo}s can be applied to the two-dimensional problem by Trotterizing over each axis direction allowing one to simulate models with interactions between collinear sites, i.e. between sites with either the same $x$ or $y$ lattice coordinate.
These so-called axial interactions, occurring between collinear sites on the lattice, are present in the axial~\cite{fisher_infinitely_1980, selke_annni_1988} and biaxial next-nearest neighbor Ising models (ANNNI and BNNNI respectively) commonly studied in the context of frustrated quantum magnetism~\cite{eckstein_large-scale_2024}.
Axial long-range interactions, decaying with an inverse power law of distance have also been studied appearing in the context of classical spin lattices and the quantum-classical correspondence~\cite{defenu_anisotropic_2016}.

Including interactions between diagonal sites is trickier, requiring a more sophisticated treatment.
Next-nearest neighbor (NNN) terms have been 
accounted for in the finite temperature context, using the so-called next-nearest neighbor update, first considered with application to fermions~\cite{corboz_simulation_2010} and later to the classical $J_1$-$J_2$ Heisenberg model~\cite{gauthe_thermal_2022, gauthe_thermal_2023}, where this method has been associated with problematic, fictitious breaking of $C_{4v}$ symmetries.
In addition, extending beyond NNN interactions requires increasing unit-cell sizes, quickly becoming impractical.

As for long-range interactions, one proposal couples a finite-state machine PEPO to the partition function of a statistical physics model with a two-point correlation function decaying algebraically with Euclidean distance~\cite{orourke_efficient_2018,li_generalization_2019}. However, an extension to time-evolution appears problematic at first glance as it is not obvious how $n$-point correlation functions can be turned into the required products of two-point correlation functions.
Ref.~\cite{orourke_simplified_2020} introduce the generalized matrix product operator (gMPO) capable of representing two-dimensional long-range interactions by fitting to Gaussian basis of functions. This method was subsequently applied to the problem of finding the ground state phase diagram of a Rydberg atom array~\cite{orourke_entanglement_2023}, however constructing gMPO forms of time-evolution operators, if possible, is likely represent a significant degree of technical work.


\section{\label{sec:op}Construction of the cluster time-evolution operator}
Generally, the goal is to simulate the real or imaginary time evolution of quantum many-body systems defined on a two-dimensional lattice $\Lambda$ who's mixed-state $\rho(t)$ at time $t$ in the Hilbert space $\mathcal{H}$ is represented by a tensor network ansatz.
We will consider the specific case where the time-evolution is generated by an operator $\mathcal{G}$ such that evolution from an initial state $\rho(0)$ is given by a dynamical map of the form:
\begin{equation}\label{eq:dyn-map-general}
    e^{\tau \mathcal{G}}: \rho(0) \mapsto \rho(\tau).
\end{equation}
of which Lindblad time evolution~\eqref{eq:dyn-map} is a special case.

It remains to construct efficient tensor network operator approximations to~\eqref{eq:dyn-map-general}.
We consider specifically a subset of generators $\mathcal{G}$ that can be written in the following form:
\begin{equation}\label{eq:valid-form}
	\mathcal{G} = \sum_{v_1,v_2 \in \Lambda} \left( \sum_{k = 1}^{\kmax} 
        g^{[k]}_{v_1 v_2}  \right) +  \sum_{v\in\Lambda}g^{[0]}_{v} 
,\end{equation}
where the notation $\hat{X}_{v_1\ldots v_n}$ denotes an $n$-body operator on $\mathcal{H}$ acting as $\hat{X}$ on the subspace of lattice sites $v_1,\ldots,v_n \in \Lambda$, and the identity elsewhere, i.e. \eqref{eq:valid-form} describes operators formed as sums of at most $2$-body terms.

Our goal to find an approximation to the time-evolution operator~\eqref{eq:dyn-map-general} that has an efficient representation as a tensor network. 
To simplify the explanation, we will consider the case where the generator $\mathcal{G}$ is the Hamiltonian $H$ of a quantum system, i.e. $\mathcal{G} \coloneq H$ in~\eqref{eq:dyn-map-general} with $\tau \coloneq -i t$ for real time evolution to a time $t$. 
The extension to Lindblad time evolution is deferred to \secref{sec:app}.

\subsection{Extensive Expansion}

A naive finite order expansion of the time evolution operator \eqref{eq:dyn-map-general} in increasing powers of $H$ results in an error intensive in the size of the system $N$, i.e. the number of neglected terms in the expansion \emph{per site} depends on the system size.
However, by including all terms of higher order (up to infinite order) that do not overlap, the neglected terms per lattice site becomes independent of the system size.
Specifically, Let $H^{(j)}$ be a local term in a Hamiltonian $H$ with support on a finite subset of the lattice $\lambda_j \subset \Lambda$, then the tuples of indices
$I_{n}= \left\{ j_1,j_2,\dots,j_{n} \in \Lambda \mid \lambda_i \cap \lambda_j \cap \cdots = \varnothing \right\}$
define tuples of $n$ terms in the Hamiltonian whose support do not overlap.
The decomposition:
\begin{equation}\label{eq:extensive-approx}
	e^{\tau H} \approx 1 + \tau \sum_{\mathclap{(j)\in j_{1}}} H^{(j)} +
	\tau^2 \sum_{\mathclap{(j,k)\in j_2}} H^{(j)} H^{(k)}
	+ \dots
	,\end{equation}
is then size extensive and thus well defined in the thermodynamic limit~\cite{zaletel_time-evolving_2015}, having $\mathcal{O}(N \tau^2)$ neglected terms.
For certain one-dimensional Hamiltonians, the expansion~\eqref{eq:extensive-approx} can be represented as a \ac{mpo} by appealing to the \ac{fsm} picture of the Hamiltonian~\cite{crosswhite_finite_2008, crosswhite_applying_2008}, and performing the following `loopback' procedure~\cite{vandamme_efficient_2024}: each \ac{fsm} path from the initial state to the final state generates a single term in the Hamiltonian; to generate \emph{products} of terms, one should map the final state of the \ac{fsm} back to the initial state. 
This then allows an additional term to be generated in the product-of-terms that itself forms a term in the expansion~\eqref{eq:extensive-approx}.
An $n$-th order term in~\eqref{eq:extensive-approx} is formed from $n$ repetitions of the \ac{fsm} assuming one multiplies by a factor of $\tau$ at each repetition. 

This loopback procedure generates what is referred to as the first-order \WI approximation, corresponding to the expansion~\eqref{eq:extensive-approx}.
It can be modified further to also include all second-order terms that overlap exactly once~\cite{vandamme_efficient_2024}, resulting in an improved but still first-order operator known as the \WII approximation. 
For details on how to generate such time-evolution operators in \ac{mpo} form, as well as analogous operators of arbitrary order, we refer to Ref.~\cite{vandamme_efficient_2024} and the algorithms therein. 

\subsection{Finite Signaling Agents}

\begin{table}
	\renewcommand{\arraystretch}{1.5}
	\centering
	\begin{ruledtabular}
		\begin{tabular}{c  c  c  c}
			Rule                  & \makecell{Index                                \\ $(e,s,w,n)$} & Value & Rule type \\
			\midrule
			$\fsmbox{0}{0}{0}{0}$ & $(0,0,0,0)$     & $\I$              & identity \\
			$\fsmdot{0}{0}{0}{0}$ & $(0,0,0,0)$     & $-i \hat{D}$         & D        \\
			\midrule
			$\fsmbox{1}{0}{0}{0}$ & $(1,0,0,0)$     & $J_1 \hat{C}$     & C        \\
			$\fsmbox{0}{0}{0}{1}$ & $(0,0,0,1)$     & $J_1 \hat{C}$     & C        \\
			$\fsmbox{0}{0}{1}{0}$ & $(0,0,1,0)$     & $-i \hat{B}$      & B        \\
			$\fsmbox{0}{1}{0}{0}$ & $(0,1,0,0)$     & $-i \hat{B}$      & B        \\
			\midrule
			$\fsmbox{1}{1}{0}{0}$ & $(1,1,0,0)$     & $J_2 J_1^{-1} \I$ & A        \\
			$\fsmbox{0}{1}{2}{0}$ & $(0,1,2,0)$     & $J_2 J_1^{-1} \I$ & A        \\
			$\fsmbox{2}{0}{0}{0}$ & $(2,0,0,0)$     & $-i \hat{B}$      & B        \\
		\end{tabular}
	\end{ruledtabular}
	\caption{Example of non-zero \ac{fsa} rules for the real time-evolution of the $J_1$-$J_2$ Ising-type Hamiltonian~\eqref{eq:ham-j1j2}, excluding the Hermitian conjugate ($\mathrm{h.c.}$) terms.
    Including the rules obtained via swapping $\hat{C}$ and $\hat{B}$ in the listed C- and B-type rules gives the $\mathrm{h.c.}$ terms. 
		The bond dimension of the resulting \ac{cpepo} operator is $\eta_{h} = 5$ on the horizontal direction and $\eta_{v} = 3$ on the vertical direction, or $\eta_h = 3$ and $\eta_v = 2$ for the special case of $\hat{C}=\hat{B}$. After including the correct prefactor listed in \tabref{tab:prefactor} depending on whether one wishes to generate the \WI or \WII approximation, the operator generated by the \ac{cpepo} is the sum over all the valid tilings of the listed rules.
        Note, this specific combination of rules is not unique. 
		\label{tab:fsm-j1j2}}
\end{table}

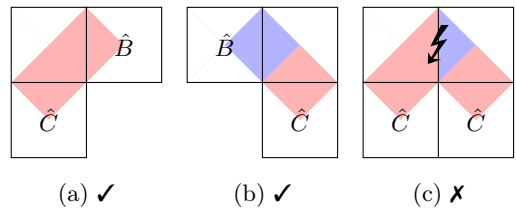
\begin{figure}
	\begin{equation*}
		\begin{tikzpicture}
			\node[diagonal fill={white}{white}{white}{red!30}{$\hat{C}$}] at (0,0) {};
			\node[diagonal fill={red!30}{red!30}{white}{white}{}] at (0,1) {};
			\node[diagonal fill={white}{white}{red!30}{white}{$\hat{B}$}] at (1,1) {};
			\node at (0.5,-1) {(a) \cmark};
		\end{tikzpicture}
		\quad
		\begin{tikzpicture}
			\node[diagonal fill={white}{white}{white}{red!30}{$\hat{C}$}] at (0,0) {};
			\node[diagonal fill={white}{red!30}{blue!30}{white}{}] at (0,1) {};
			\node[diagonal fill={blue!30}{white}{white}{white}{$\hat{B}$}] at (-1,1) {};
			\node at (-0.5,-1) {(b) \cmark};
		\end{tikzpicture}
		\quad
		\begin{tikzpicture}
			\node[diagonal fill={white}{white}{white}{red!30}{$\hat{C}$}] at (0,0) {};
			\node[diagonal fill={white}{red!30}{blue!30}{white}{}] at (0,1) {};
			\node[diagonal fill={red!30}{red!30}{white}{white}{}] at (-1,1) {};
			\node[diagonal fill={white}{white}{white}{red!30}{$\hat{C}$}] at (-1,0) {};
			\node at (-0.5,1) {\huge\Lightning};
			\node at (-0.5,-1) {(c) \xmark};
		\end{tikzpicture}
	\end{equation*}
	\caption{Examples of two terms (a) and (b) in the Hamiltonian~\eqref{eq:ham-j1j2} that are accepted by the \ac{fsa} and another term (c) not in~\eqref{eq:ham-j1j2} that is correctly rejected due to the additional (blue) level. This term would incorrectly be included in the sum had the blue level been replaced by the red level.\label{fig:unwanted}}
\end{figure}

\begin{table}[t]
	\centering
	\begin{ruledtabular}
		\begin{tabular}[c]{c  c c  c  c  c}
			           &               & \multicolumn{4}{c}{\WII}                                                     \\
			\cmidrule{3-6}
			Rule type  & \WI           & id                       & A      & B $\lor$ C       & D                     \\
			\midrule
			id         & $1$           & $1$                      & $1/2$  & $\sqrt{\tau}/2$  & $\tau/2$              \\
			A          & $1$           &                          & $1/2!$ & $\sqrt{\tau}/2!$ & $\tau/2!$             \\
			B $\lor$ C & $\sqrt{\tau}$ &                          &        & $\tau/2!$        & $\tau\sqrt{\tau} /2!$ \\
			D          & $\tau$        &                          &        &                  & $\tau^2 /2!$
		\end{tabular}
	\end{ruledtabular}
	\caption{Prefactor required when constructing products of intersecting rules based on rule type. The table is symmetric. This combination of prefactors is not unique, but is convenient to avoid breaking any symmetry between lattice directions.}
	\label{tab:prefactor}
\end{table}

We now describe how to construct the \WI and \WII operators in two-dimensions in terms of the \ac{fsa} generalization of the \ac{fsm} picture.
Suppose for simplicity that $\Lambda$ is a two-dimensional square lattice.
Let $\mathcal{H}$ be the Hilbert space of a many-body quantum system defined on $\Lambda$.
Now associate to each edge in the lattice an integer, denoted a \emph{signal}, such that for each vertex $v$ there is a tuple of signals, $(e,s,w,n)$ where $e,s,w,n \in \{0,...,l-1\}$, defining the signal on each of the $4$ edges incident to $v$ (denoted here using the cardinal directions).
Let $\hat{O}$ be an operator, acting on $\mathcal{H}$, of the form
\begin{equation}
	\hat{O} = \bigotimes_{v \in \Lambda} \hat{X}^{[v]}
	\label{eq:fsa-term}
\end{equation}
such that for all $v \in \Lambda$ the operator $\hat{X}^{[v]}$ is determined by the specific combination of signals on the edges incident to $v$.
If one constructs the operator-valued tensor $\hat{X}_{eswn}$ (where for simplicity we have imposed translational invariance $\hat{X}^{[v]}_{eswn} = \hat{X}_{eswn} \, \forall v \in \Lambda$), then the corresponding projected entangled pair operator generated by $\hat{X}_{eswn}$ is precisely the sum of the operators $\hat{O}$ that would result from every possibly combination of signals assigned to all the edges in the lattice.

This is largely trivial until we impose that $\hat{X}_{eswn} = \hat{0}$ for some combinations of $(e,s,w,n)$.
If any of the vertices receives such a combination, then immediately $\hat{O} = \hat{0}$, and we say the term $\hat{O}$ is \emph{rejected}.
All it takes is one such operator in the tensor product~\eqref{eq:fsa-term} to force the term to be rejected.
We say that the map $(e,s,w,n) \to \hat{X}$ defines a \ac{fsa} \emph{rule}.
By carefully choosing the non-zero rules, or in other words, elements of the tensor $\hat{X}_{eswn}$, one can engineer the resulting sum of non-rejected terms to coincide with the cluster expansion~\eqref{eq:extensive-approx} of the desired Hamiltonian, provided we make sure that the rules also accommodate for the correct prefactor of $\tau^{n}$, which will be described shortly.

Note, at a minimum one must have the zero'th order term $\I$ in~\eqref{eq:extensive-approx} so no matter the Hamiltonian, one must have the rule $\hat{X}_{0000} = \I$.
This is denoted the background or \emph{identity} rule.
We also require that all accepted $\hat{O}$ represent a  valid product of complete terms in $H$.
To enforce this, it is useful to place each rule in one of four categories: C-type rules define the start, or \emph{head} of a given term in $H$; similarly, B-type rules define the end, or \emph{tail};
rules of A-type define \emph{bulk} rules that are part of a term but are neither the head nor the tail; the single D-type rule is equal to the sum over all the one-local terms in $H$.
Using this categorization, each term in $H$ should (a) either by formed from one C-type rule, one B-type rule, and any number of (possibly zero) A-type rules, or (b) be a one-body term included in the single D-type rule.

Once these rules are defined and categorized, it remains to resolve for the pre-factor of $\tau^n$ depending on how many terms in $H$ appear in each term in $\hat{O}$.
One solution is to modify the rules such that each term in $H$ gains a factor $\tau$.
This can be achieved by multiplying either the C-type rule \emph{or} the B-type rule by $\tau$ or, as shown in \tabref{tab:prefactor}, both rules by $\sqrt{\tau}$. 
This defines the two-dimensional generalization of the \WI approximation.
\begin{algorithm}[t]
	\hrulefill
	\vspace{-8pt}
	\hrulefill
	\begin{algorithmic}[1]
		\Require Hamiltonian $H$ in the form of FSA rules stored as the rank-$z$ tensor $R$, where $z$ is the coordination number of the lattice.
		\Ensure Tensor $P$ who's network defines the \WII approximation to the time evolution generated by $H$.
		\Procedure{IncludeOverlapping}{$R$}
		\State $I_R \gets$ indices of $R$
		\State $P \gets $ tensor of zeros with indices $I_R$
		\For{$a \in I_R, b \in I_R$}
		\If{\textbf{any} $a[n] \neq 0 \land b[n] \neq 0$ \textbf{for} $n \in [1,z]$}
		\State \textbf{skip}
		\Else
		\State $c \gets a + b$
		\State $\tau \gets$ prefactor of $R[a] \times R[b]$ \Comment{From~\tabref{tab:prefactor}}
		\State $P[c] \gets P[c] + \tau R[a] R[b] $
		\EndIf
		\EndFor
		\State \textbf{return} $P$
		\EndProcedure
	\end{algorithmic}
	\hrulefill
	\vspace{-8pt}
	\hrulefill
	\caption{Generate the \WII approximation to a Hamiltonian defined in terms of FSA rules.}\label{alg:wii}
\end{algorithm}
At no increase of the number of signals required, the \WII approximation can be generated by considering products of terms who's support intersects only once.
This manifests as allowing for products of rules that do not have any of the same signal non-zero.
The background $0$ signal essentially carries no information, and can be used to store the signals from the second operator in the product.
If non-zero rules do overlap, then additional signals are needed (increasing the subsequent bond dimension of the tensor network operator) and are therefore ignored.
The correct \emph{second order} prefactor which can be determined from \tabref{tab:prefactor}, and the tensor network operator corresponding to the \WII approximation can be obtained from the FSA rules using Algorithm~\ref{alg:wii}.

\subsubsection{Example: $J_1$-$J_2$ Hamiltonian}

It is useful to consider an example.
Consider the following $J_1$-$J_2$ Ising-like next-nearest neighbor Hamiltonian:
\begin{equation}\label{eq:ham-j1j2}
	H_{\mathrm{nnn}} =
	\left(J_1 \sum_{\mathclap{\la j, k \ra 
    \in \Lambda}}\hat{C}_j \hat{B}_k + J_2 \sum_{\mathclap{\la\!\la j, k \ra\!\ra \in \Lambda}} \hat{C}_j \hat{B}_k  + \mathrm{h.c.} \right) + \sum_{j\in\Lambda} \hat{D}_j,
\end{equation}
where the first sum is over the nearest-neighbors and the second sum is over the next-nearest neighbors (those directly diagonal to each other). We use $\mathrm{h.c.}$ as an abbreviation for the Hermitian conjugate of the written terms and make the choice $\hat{B} = \hat{C}^{\dagger}$ for all sites in the lattice.

The basic rules defining the \ac{fsa} representation of this Hamiltonian are listed in \tabref{tab:fsm-j1j2}, \emph{without} prefactors of $\tau$. 
The resulting bond dimension of the tensor network operator is then, excluding the $\mathrm{h.c.}$ terms, $\eta_{h} = 3$ on the horizontal lattice direction, and $\eta_{v} = 2$, where the so-called unconventional path, indexed by $2$, is not required. 
The total bond dimension including the $\mathrm{h.c.}$ terms is then $\eta_{h} = 5$ and $\eta_{v} = 3$.
The un-conventional path is needed to prevent the \ac{fsa} generating terms not in the Hamiltonian, as shown in~\figref{fig:unwanted}.
Note, this is not the only combination of rules capable of generating $H_{\mathrm{nnn}}$.
An alternative set of rules that lead to a symmetric tensor network operator for the common case of $\hat{C}_{j} \hat{B}_{k} = \hat{X}_j\hat{X}_k $ for all $ j,k \in \Lambda$ is listed in Appendix.~\ref{app:extra}.
%

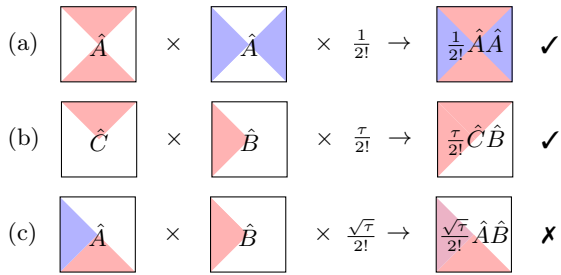
\begin{figure}
	\begin{align*}
		 & \begin{tikzpicture}
			   \node at (-1,0) {(a)};
			   \node[diagonal fill={white}{red!30}{white}{red!30}{$\hat{A}$}] at (0,0) {};
			   \node at (1,0) {$\times$};
			   \node[diagonal fill={blue!30}{white}{blue!30}{white}{$\hat{A}$}] at (2,0) {};
			   \node at (3,0) {$\times$};
			   \node at (3.5,0) {$\frac{1}{2!}$};
			   \node at (4,0) {$\to$};
			   \node[diagonal fill={blue!30}{red!30}{blue!30}{red!30}{$\frac{1}{2!}\hat{A}\hat{A}$}] at (5,0) {};
			   \node at (6,0) {\cmark};
		   \end{tikzpicture} \\
		 & \begin{tikzpicture}
			   \node at (-1,0) {(b)};
			   \node[diagonal fill={white}{white}{white}{red!30}{$\hat{C}$}] at (0,0) {};
			   \node at (1,0) {$\times$};
			   \node[diagonal fill={white}{white}{red!30}{white}{$\hat{B}$}] at (2,0) {};
			   \node at (3,0) {$\times$};
			   \node at (3.5,0) {$\frac{\tau}{2!}$};
			   \node at (4,0) {$\to$};
			   \node[diagonal fill={white}{white}{red!30}{red!30}{$\frac{\tau}{2!}\hat{C}\hat{B}$}] at (5,0) {};
			   \node at (6,0) {\cmark};
		   \end{tikzpicture} \\
		 & \begin{tikzpicture}
			   \node at (-1,0) {(c)};
			   \node[diagonal fill={white}{red!30}{blue!30}{white}{$\hat{A}$}] at (0,0) {};
			   \node at (1,0) {$\times$};
			   \node[diagonal fill={white}{white}{red!30}{white}{$\hat{B}$}] at (2,0) {};
			   \node at (3,0) {$\times$};
			   \node at (3.5,0) {$\frac{\sqrt{\tau}}{2!}$};
			   \node at (4,0) {$\to$};
			   \node[diagonal fill={white}{red!30}{purple!30}{white}{$\frac{\sqrt{\tau}}{2!} \hat{A}\hat{B}$}] at (5,0) {};
			   \node at (6,0) {\xmark};
		   \end{tikzpicture}
	\end{align*}
	\caption{(a) and (b) example of products of \ac{fsa} rules that do not require additional levels, and can therefore be included at no cost to the operator bond dimension. In (c), the product includes overlapping non-trivial levels so we require an additional level to transmit both the blue and red signals simultaneously.
		Including such product rules allows the \ac{fsa} to also generate all the second order terms that intersect only once.}
\end{figure}

\section{\label{sec:lr}Long-range interactions}
We now describe how long-range interactions can be represented in two-dimensions using \ac{fsa}.
By long-range we mean operators that act on arbitrary pairs of lattice sites at any distance from each other.
Similar to the \ac{fsm} construction for \ac{mpo}s~\cite{crosswhite_applying_2008, zaletel_time-evolving_2015, vandamme_efficient_2024}, the \ac{fsa} construction allows us to represent Hamiltonians with long-range interactions by allowing the machine to remain on a given state indefinitely.
Suppose we have a Hamiltonian $H$ on a lattice $\Lambda$ of the form
\begin{equation}
	H = \left(\sum_{j \neq k \in \Lambda} V(x,y) \hat{C}_{j}\hat{B}_{k} + \mathrm{h.c.}  \right)+\sum_{j\in \Lambda} \hat{D}_{j}
	\label{eq:target-ham}
	,\end{equation}
where $x$ and $y$ is the distance between the two lattice sites $j$ and $k$ along each respective lattice direction.
In addition to rules for the $J_1$-$J_2$ Hamiltonian listed in \tabref{tab:fsm-j1j2}, introducing the following A-type rules:
\begin{equation}\label{eq:rules-all-to-all}
	\fsmbox{1}{0}{1}{0} =
	\fsmbox{0}{1}{0}{1} =
	\fsmbox{2}{0}{2}{0} = \lambda \I
\end{equation}
and also setting $J_2 J_1^{-1} = \lambda$ generates terms acting between lattice sites at all distances.
The specific case of $\lambda = 1$ corresponds to all-to-all interactions, $V(x,y) = 1$, however this results in diverging interactions in the thermodynamic limit.
The other limit $\lambda = 0$ is defines a model with completely localized interactions.

\subsection{Gaussian approximation to the power law}
\begin{table}[t]
	\setlength{\tabcolsep}{5pt}
	\centering
	\begin{ruledtabular}
		\begin{tabular}[c]{c  c  c}
			\makecell{Index                                              \\
			$(e,s,w,n)$}  & Value                            & Rule type \\
			\midrule
			$(0,0,0,0)$   & $\I$                             & identity  \\
			$(0,0,0,0)$   & $ -i \hat{D}/{\kmax}$            & D         \\
			\midrule
			\multicolumn{3}{l}{For all $i$ and $j$ in $1$ to $n_{\max}$} \\
			\midrule
			$(i,0,0,0)$   & $-i t_k \hat{C}$                 & C         \\
			$(0,0,0,i)$   & $-i t_k \hat{C}$                 & C         \\
			$(0,0,i,0)$   & $ s_{k,i} \lambda_{k,i} \hat{B}$ & B         \\
			$(0,i,0,0)$   & $ s_{k,i} \lambda_{k,i} \hat{B}$ & B         \\
			$(i,0,i,0)$   & $\lambda_{k,i}\I$                & A         \\
			$(0,i,0,i)$   & $\lambda_{k,i}\I$                & A         \\
			$(i',0,i',0)$ & $\lambda_{k,i} \I$               & A         \\
			$(j,i,0,0)$   & $ s_{k,j}\lambda_{k,j}\I$        & A         \\
			$(0,i,j',0)$  & $s_{k,j}\lambda_{k,j}\I$         & A         \\
		\end{tabular}
	\end{ruledtabular}
	\caption{Example of non-zero \ac{fsa} rules that generate an approximation to a radial Gaussian function. The primed levels $i'$ correspond to the so-called unconventional levels and should be given a distinct index to the unprimed level $i$.
    These rules should be used in combination with~\tabref{tab:prefactor} and Algorithm~\ref{alg:wii}.
    }
	\label{tab:fsm-long-range}
\end{table}

When $\abs{\lambda} < 0$, the strength of the interaction between two sites decays exponentially as a function of the Manhattan distance, $V(x,y) = \lambda^{\abs{x}+\abs{y}}$.
Functions of the Euclidean distance $r = \sqrt{x^2 + y^2}$ can be constructed using a \ac{fsa} by first approximating a Gaussian along each lattice direction as a finite sum of exponential functions~\cite{orourke_simplified_2020},
$
	e^{-\mu x^2} \approx g_{\mu}(x) = \sum^{n_k}_{k = 1} s_k \lambda_k^{x}
	,$
such that the \ac{fsa} generates the product $g_{\mu}(x)g_{\mu}(y) \approx g_{\mu}(r)$.
Then, defining a set of $\kmax$ radial basis functions
$f_k(x,y) = g_{\mu_k}(x)g_{\mu_k}(y)$,
the decay profile $V(x,y)$ can be approximated by the weighted sum $V(x,y) \approx \sum_{k = 1}^{\kmax} t_k f_k(x,y)$ for $r = \sqrt{x^2 + y^2} > 0 $.
Note, for anisotropic lattices, one can have $g_{\mu_k}(x) \neq g_{\nu_{k}}(y)$ in general, however we only consider the simpler isotropic case.

\subsubsection{Computing the optimal weights}
The decay profile $V(x,y)$ is approximated by a function $f_{\mathrm{fit}}(x,y)$ defined in terms of the parameters,
\begin{equation}
	P_{f} = \{t_k, s_1, \ldots, s_{n_{k}}, \lambda_{1}, \ldots, \lambda_{n_{k}} \}_{k\in 1 \isep \kmax}
	.\end{equation}
we then wish to solve for $P_{f}$ the optimization problem
\begin{equation}
	\min_{P_f}  \norm{V(x,y) - f_{\mathrm{fit}}(x,y)}
	\label{eq:min-problem}
\end{equation}
on the disc $D_{r_{\mathrm{cut}}} = \{0 < \sqrt{x^2 + y^2} < r_{\mathrm{cut}} \mid x,y \in \mathbb{Z}\}$, where $r_{\mathrm{cut}}$ is some finite radius cutoff.
To do so, we use the algorithm described in Ref.~\cite{pirvu_matrix_2010} to compute the optimal expansion of the function $V(x,y)$ in terms of Gaussians.
For $V(x,y) = r^{\alpha}$, we can directly apply the algorithm in Ref.~\cite{pirvu_matrix_2010} by finding the solution to \eqref{eq:min-problem} using $V(x,y) = r^{\alpha/2}$ in terms of exponentials $\lambda_k^{r}$ and then subsequently taking $r \to r^2$ to find an approximation to $r^{\alpha}$ in terms of Gaussians $g_{\mu_k}(r) = e^{-\mu_k r^2}$
Then, assuming isotropy between the lattice directions, we again solve
with $\mu_k = -\log \lambda_k$ where it is useful to allow for $s_k, \lambda_k \in \mathbb{C}$.
Note, $g_{\mu_k}(r)$ only needs to be a good approximation of $e^{-\mu r^2}$ for $r \in \mathbb{Z}$ in contrast to $f_{\mathrm{fit}}$ where the elements of $D_{r_{\mathrm{cut}}}$ are non-integer in general.
\subsubsection{Quality of approximation}

The parameters that control the quality of approximation are the number of Gaussian, $\kmax$ that approximate the target function and, for each $k$, the number of exponential functions used to approximate each Gaussian $g_{\mu_k}(r)$ denoted $n_{k,\max}$. 
Depending on the value of $\mu_k$, simply increasing $n_{k,\max}$ may not improve the approximation and may make it worse. 
As such, $n_{k,\max}$ is determined by choosing an integer $n_{\max}$ such that $n_{k,\max} \leq n_{\max}$ and then selecting $n_{k,\max}$ that gives the best approximation for each $k$.
An additional parameter $g_{\tol}$ defines the allowed tolerance in error $\epsilon_{\mu_k} =\sum_{r = 1}^{N}\abs*{g_{\mu_k}(r) - e^{\mu_k r^2} }$ where an approximation with $n_{k,\max} < n_{\max}$ is selected. 
Any given value of $g_{\tol}$ then results in a non-unique combination of $n_{k,\max}$ for a given $\kmax$. 

\subsubsection{Time evolution operator}
We can approximate the time evolution generated by the target Hamiltonian $H$ given in~\eqref{eq:target-ham} by first defining
$
	H^{[k]} = t_k \sum_{i, j \in \Lambda}  f_k(x,y) \hat{C}_{i} \hat{B}_{j} + \sum_{i \in \Lambda} \hat{D}_i/\kmax
	.$
We can now define the approximated fitting Hamiltonian $H_{\mathrm{fit}}$ as
$
	H \approx H_{\mathrm{fit}} = \sum_{k  = 1}^{\kmax}  H^{(k)}
$
provided $f_{\mathrm{fit}}(x,y) \approx V(x,y)$.
The Hamiltonian $H_{\mathrm{fit}}$ can then be represented as a \ac{fsa} with $\mathcal{O}(\kmax n_{\max})$ non-trivial levels.
This can be reduced to $\mathcal{O}(n_{\max})$ levels when constructing the time evolution operator by taking the Suzuki-Trotter decomposition~\cite{trotter_product_1959, suzuki_generalized_1976},
\begin{equation}\label{eq:trotter}
	e^{ \tau H_{\mathrm{fit}}} =
	W^{[1]}(\tau)
	W^{[2]}(\tau)
	\dots
	W^{[\kmax]}(\tau)
	+ \mathcal{O}(\tau^2)
	,\end{equation}
where
$
	W^{[k]}(\tau) = \exp(\tau H^{[k]})
	\label{eq:wk-op}
	,$
at the cost of $\kmax$ update and truncation steps per time step.
In terms of the size extensive cluster expansion~\eqref{eq:extensive-approx}, each $\Wk$ can then be represented by tensor network operator generated by the \ac{fsa} rules listed in \tabref{tab:fsm-long-range} each with bond dimension $\eta_h = 1 + 2  n_{k, \max}$ along the horizontal direction and $\eta_v = 1 + n_{k, \max}$ along the vertical bonds where we do not requiring the additional level corresponding to the unconventional path.
A vectorized Liouvillian super-operator representation would require (assuming one-local Lindblad operators) a real bond dimension of $\eta_k = (1 + 4 n_{k,\max}, 1 + 2 n_{k,\max})$ as one effectively has two entirely independent terms in the Hamiltonian, one from the ket space and one from the bra space.

One could in principle perform a Suzuki-Trotter decomposition over the terms appearing in $H^{(k)}$ that define the approximation to the Gaussian, however it was found that doing so resulted in instabilities caused by terms with very large interaction coefficient that would ordinarily cancel out when summed by the \ac{fsa}.

\section{\label{sec:app}Application to Lindblad Time Evolution}
We now describe how the operator construction just described can be applied to the case of Lindblad time evolution.
\subsection{Tensor Network State Ansatz}

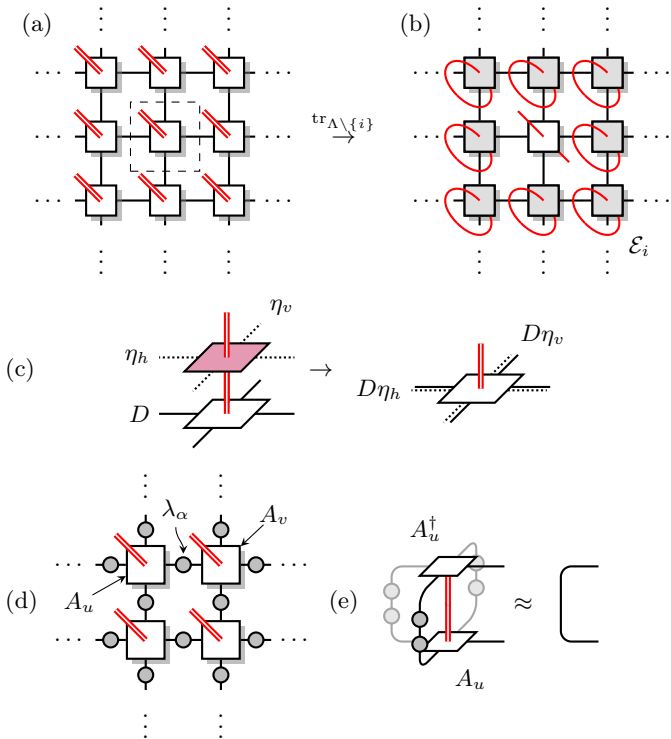
\begin{figure}
	\centering
	\begin{flalign*}
		\vcenter{\hbox{
				\begin{tikzpicture}[scale=0.85]
					\foreach \y in {1,2,3}{
							\node (0\y) at (0.2,\y) {$\dots$};
							\node (4\y) at (3.8,\y) {$\dots$};
							\draw[thick] (0\y) to (4\y);
						}
					\foreach \x in {1,2,3}{
							\node (\x 0) at (\x,0.1) {$\strut \vdots$};
							\node (\x 4) at (\x,3.9) {$\strut \vdots$};
							\draw[thick] (\x 0) to (\x 4);
							\foreach \y in {1,2,3}{
									\node[draw, thick, fill=white, minimum size=0.4cm, drop shadow](\x\y) at (\x,\y) {};
									\draw[thick, red, double] (\x\y.center) --++ (-0.4,0.4);
								}
						}
					\node[draw, dashed, fit = (22), inner sep=7]{};
                    \node at (0,3.75) {(a)};
				\end{tikzpicture}}}
		\overset{\tr_{\Lambda \setminus \{i \} }}{\to}
		\vcenter{\hbox{
				\begin{tikzpicture}[scale=0.85]
					\begin{scope}[on background layer]
						\foreach \x in{1,2,3}{
								\foreach \y in {1,2,3}{
										\coordinate (\x\y a) at (\x,\y);
										\coordinate[xshift=-0.3cm,yshift=-0.3cm] (\x\y b) at (\x\y a);
										\node[draw, thick, fill=white, minimum size=0.4cm, drop shadow](\x\y) at (\x\y a) {};
									}
							}
						\foreach \y in {1,2,3}{
								\node (0\y) at (0.2,\y) {$\dots$};
								\node (4\y) at (3.8,\y) {$\dots$};
								\foreach \x [count=\xp] in {0,1,2,3}{
										\draw[thick] (\x\y) to (\xp\y);
									}
							}
						\foreach \x in {1,2,3}{
								\node (\x 0) at (\x,0.1) {$\strut \vdots$};
								\node (\x 4) at (\x,3.9) {$\strut \vdots$};
								\foreach \y [count=\yp] in {0,1,2,3}{
										\draw[thick] (\x\y) to (\x\yp);
									}
							}
					\end{scope}
					\foreach \x in {1,2,3}{
							\foreach \y in {1,2,3}{
									\ifthenelse{ \x = 2 \and \y = 2}{
										\draw[thick, red] (\x\y a) --++ (0.4, -0.4);
										\node[draw, thick, fill=white, minimum size=0.4cm] at (\x\y a) {};
										\draw[thick, red] (\x\y a) --++ (-0.4, 0.4);
									}
									{
										\draw[thick, red] (\x\y b) to [out=-45, in=-45, looseness=3] (\x\y a);
										\node[draw, thick, fill=lightgray!50, minimum size=0.4cm] at (\x\y a) {};
										\draw[thick, red] (\x\y a) to [out=135, in=135, looseness=3] (\x\y b);
									}
								}
						}
					\node[above] at (3.5,0) {$\mathcal{E}_i$};
                    \node at (0,3.75) {(b)};
				\end{tikzpicture}}}&&
	\end{flalign*}
    \vspace{-2em}
    	\begin{flalign*}
        \mathrm{(c)}\quad\quad\quad	
        \vcenter{\hbox{
				\begin{tikzpicture}[x={(1.5cm,0cm)},y={(0.75cm,0.75cm)},z={(0cm,1.5cm)}]
					\draw[thick] (0,0,0) -- (0,+0.6,0);
					\draw[thick] (0,0,0) -- (0,-0.6,0);
					\draw[thick] (0,0,0) -- (+0.6,0,0);
					\draw[thick] (0,0,0) -- (-0.6,0,0) node[left] {$D$};
					\draw[thick, fill=white] (0 - 0.25, 0 - 0.25, 0) -- ++(0.5, 0, 0) -- ++(0,0.5,0) -- ++(-0.5,0,0) -- cycle;
					\draw[thick, red, double] (0, 0, 0) -- ++(0,0,0.4);
					\draw[thick, densely dotted] (0,0,0.5) -- ++ (0,+0.6,0) node[above right] {$\eta_{v}$};
					\draw[thick, densely dotted] (0,0,0.5) -- ++ (0,-0.6,0);
					\draw[thick, densely dotted] (0,0,0.5) -- ++ (+0.6,0,0);
					\draw[thick, densely dotted] (0,0,0.5) -- ++ (-0.6,0,0) node[left] {$\eta_{h}$};
					\draw[thick, red, double] (0, 0, 0.5) -- ++(0,0,-0.4);
					\draw[thick, fill=purple!40] (0 - 0.25, 0 - 0.25, 0.5) -- ++(0.5, 0, 0) -- ++(0,0.5,0) -- ++(-0.5,0,0) -- cycle;
					\draw[thick, red, double] (0, 0, 0.5) -- ++(0,0,0.4);
				\end{tikzpicture}}}
		\rightarrow
		\vcenter{\hbox{
				\begin{tikzpicture}[x={(1.5cm,0cm)},y={(0.75cm,0.75cm)},z={(0cm,1.5cm)}]
					\draw[thick] (0.03,0,0) --++ (0,+0.6,0);
					\draw[thick] (0.03,0,0) --++ (0,-0.6,0);
					\draw[thick] (0,0.03,0) --++ (+0.6,0,0);
					\draw[thick] (0,0.03,0) --++ (-0.6,0,0) ;
					\draw[thick, densely dotted] (-0.03,0,0) --++ (0,+0.6,0) node[above right] {$D\eta_v $};
					\draw[thick, densely dotted] (-0.03,0,0) --++ (0,-0.6,0);
					\draw[thick, densely dotted] (0,-0.03,0) --++ (+0.6,0,0);
					\draw[thick, densely dotted] (0,-0.03,0) --++ (-0.6,0,0) node[left] {$D\eta_h $};
					\draw[thick, fill=white] (0 - 0.25, 0 - 0.25, 0) -- ++(0.5, 0, 0) -- ++(0,0.5,0) -- ++(-0.5,0,0) -- cycle;
					\draw[thick, red, double] (0, 0, 0) -- ++(0,0,0.4);
				\end{tikzpicture}}}&&
	\end{flalign*}
        \vspace{-2em}
    	\begin{flalign*}
        \textnormal{(d)}
		\vcenter{\hbox{
				\begin{tikzpicture}[bond/.style={fill=gray!50, draw, thick, circle, inner sep=2},]
					\node[draw, thick, fill=white, minimum size=0.5cm, drop shadow](11) at (1,1) {};
					\node[draw, thick, fill=white, minimum size=0.5cm, drop shadow](12) at (1,2) {};
					\node[draw, thick, fill=white, minimum size=0.5cm, drop shadow](21) at (2,1) {};
					\node[draw, thick, fill=white, minimum size=0.5cm, drop shadow](22) at (2,2) {};
					\foreach \y in {1,2}{
							\node (0\y) at (0,\y) {$\dots$};
							\node (3\y) at (3,\y) {$\dots$};
							\foreach \x [count=\xp] in {0,1,2}{
									\draw[thick] (\x\y) to node[midway, fill, inner sep=1] (\x\xp\y) {} (\xp\y);
									\node[bond] at (\x\xp\y) {};
								}
						}
					\foreach \x in {1,2}{
							\node (\x 0) at (\x,-0.1) {$\strut \vdots$};
							\node (\x 3) at (\x,3.1) {$\strut \vdots$};
							\foreach \y [count=\yp] in {0,1,2}{
									\draw[thick] (\x\y) to node[midway, fill, inner sep=1] (\x\y\yp) {} (\x\yp);
									\node[bond] at (\x\y\yp) {};
								}
							\foreach \y in {1,2}{
									\draw[thick, red, double] (\x\y.center) --++ (-0.4,0.4);
								}
						}
					\node[inner sep=4] (lampnt) at (122) {};
					\node[above, yshift=0.5cm, xshift=-0.1cm ] (lamlab) at (lampnt){$\lambda_{\alpha}$};
					\draw[-stealth] (lamlab) to [in=90, out=-90](lampnt) ;
					\node[left, xshift=-0.7cm, inner sep=-1] (ulab) at (112) {$A_u$};
					\node[right, xshift=0.5cm, yshift=0.2cm, inner sep=0] (vlab) at (223) {$A_v$};
					\draw[-stealth] (ulab) to (12);
					\draw[-stealth] (vlab) to (22);
				\end{tikzpicture}
			}}
        \textnormal{(e)}
		\vcenter{\hbox{
				\newcommand{\peps}[3]{%
					\pgfmathsetmacro{\col}{iseven(#1 + #2) ? "white": "white"}
					\draw[thick, fill=white] (#1 - 0.25, #2 - 0.25,#3) -- ++(0.5, 0, 0) -- ++(0,0.5,0) -- ++(-0.5,0,0) -- cycle;
					\draw[thick, red,double] (#1, #2, #3) -- ++(0,0,0.52);
				}
				\newcommand{\pepsconj}[3]{%
					\draw[thick, red,double] (#1, #2, #3) -- ++(0,0,-0.52);
					\draw[thick, fill=white] (#1 - 0.25, #2 - 0.25,#3) -- ++(0.5, 0, 0) -- ++(0,0.5,0) -- ++(-0.5,0,0) -- cycle;
				}
				\begin{tikzpicture}[x={(1cm,0cm)},y={(0.5cm,0.5cm)},z={(0cm,1cm)}, bond/.style={fill=gray!20, draw, thick, circle, inner sep=2},]
					\draw[rounded corners,thick,gray!70] (0,+0.25,1) --++ (0,0.5,0) --++ (0,0,-1) node[pos=0.33,bond] {} node[pos=0.66, bond] {}--++ (0,-0.5,0);
					\draw[rounded corners,thick,gray!70] (-0.25,0,1) --++ (-0.5,0,0) --++ (0,0,-1) node[pos=0.33,bond] {} node[pos=0.66, bond] {}--++ (0.5,0,0);
					\peps{0}{0}{0};
					\pepsconj{0}{0}{1};
					\draw[rounded corners,thick] (0,-0.25,1) --++ (0,-0.5,0) --++ (0,0,-1) node[pos=0.33,bond, fill=gray!50] {} node[pos=0.66,bond, fill=gray!50] {}--++ (0,0.5,0);
					\draw[thick] (0.25,0,0) --++ (0.5,0,0);
					\draw[thick] (0.25,0,1) --++ (0.5,0,0);
					\node at (-0.3,0,1.5) {$A_{u}^{\dagger}$};
					\node at (0.3,0,-0.5) {$A_{u}$};
					\draw[rounded corners,thick] (2,0,1) --++ (-0.5,0,0) --++ (0,0,-1) node[pos=0.5, xshift=-0.5cm] {$\approx$}--++ (0.5,0,0);
				\end{tikzpicture}
			}}&&
	\end{flalign*}
	\caption{(a) \ac{ipepo} parameterization of the vectorized state density matrix assuming a single-site unit-cell shown in the dashed box. The doubled red line denotes the physical bonds associated with each local vectorized Hilbert space $\mathcal{H}_v^{\dagger} \otimes \mathcal{H}_v$. (b) After un-vectorizing the state, local density matrices are obtained by computing the trace environment $\mathcal{E}_i$ with respect to the $i$-th site in the lattice using an appropriate boundary algorithm as discussed in Appendix~\ref{app:vumps}.
    (c) Application of a tensor network super-operator to a
single tensor of the vectorised iPEPO ansatz. (d) A state in the Vidal gauge has additional bond weights $\lambda_{\alpha}$ for each bond $\alpha = (u,v)$ in the lattice $\Lambda$. In the Vidal gauge, for each tensor $A_u$ in the lattice, the quasi-orthogonality condition (e) is satisfied for all bonds incident to that tensor.
    }
	\label{fig:unit-cell}
\end{figure}

We represent the density matrix on an infinite, two-dimensional square lattice as an \ac{ipepo} tensor network where each vertex $v\in\Lambda$ in the lattice is associated with a rank-$6$ tensor $A_v$ with shape $(d,d,D,D,D,D)$.
An edge between two vertices indicates a tensor contraction between the associated indices of the tensors at those vertices.
The vectorization $\vecp$ is then obtained by fusing the physical index corresponding to the bra space with that of the ket space for all $\{A_v\}$, resulting in a network of rank-$5$ tensors with shape $(d^2, D, D, D, D)$ functionally equivalent to an \ac{ipeps}.
The tensor network structure is shown diagrammatically in \figref{fig:unit-cell}a.

\subsection{Vectorized Lindblad Equation}

The Liouvillian super-operator is linear operator on the set of density matrices and can therefore be expressed as matrix of the form
\begin{multline}
	\vecc \mathcal{L} = -i \left(\I \otimes H  - H^{\top} \otimes \I \right)\\
	+ \sum_{k} L_k^{*} \otimes L_k -
	\frac{1}{2}\bb{ \I \otimes L^{\dagger}_k L_k +  L^{\top}_k L^{*}_k\otimes \I}
	\label{eq:vec-lindblad}
,\end{multline}
acting on the set of vectorized density matrices $\vecp$ defined previously. 
We then consider time-evolution generated by operators~\eqref{eq:vec-lindblad} that can be written in the form of~\eqref{eq:valid-form}, admitting a \ac{fsa} representation, as described in \secref{sec:op} and \secref{sec:lr}.
Further details of this construction can be found in Appendix~\ref{app:lindblad-cons}.

The density matrix is then evolved from a initial state at $t = 0$ to a state at time $t = n \Delta t$ via $n$ repeated applications of the sequence of operators~\eqref{eq:trotter} (with $\tau \coloneq \Delta t$) to each tensor in the \ac{ipepo} as shown in \figref{fig:unit-cell}c.
In the limit $\Delta t \to 0$ both the Suzuki-Trotter approximation~\eqref{eq:trotter} and the extensive cluster expansion~\eqref{eq:extensive-approx} become exact.

\subsection{Bond truncation}

After a single application of the \ac{cpepo}, all four bonds on each of the \ac{ipepo} tensors are simultaneously enlarged by a multiplicative factor of the \ac{cpepo} bond dimension $\eta$, i.e. $D \to D \eta$ (see \figref{fig:unit-cell}c).
To prevent the bond dimension growing exponentially, each bond is truncated back down to $D$ using the simple-update (SU) method.
SU requires putting the \ac{ipepo} in the Vidal gauge, where each bond hosts a matrix, known as the bond weight, which are used to construct a rank-1 approximation to the environment surrounding each bond~\cite{tindall_gauging_2023}.
When truncating the bond $\alpha$ from $D' > D$ back down to $D$, one replaces the $D' \times D'$ bond weight $\lambda_{\alpha}$ such that $\lambda_{\alpha}\to U \tilde{\lambda}_{\alpha} V^{\dagger}$, where $\tilde{\lambda}$ is a $D \times D $ matrix, and $U_{\alpha}$ and $V_{\alpha}$ are $D'\times D$ rectangular isometries.
This truncates the dimension of the bond to $D$.
How $U_{\alpha}$ and $V_{\alpha}$ are obtained is described in Appendix~\ref{app:su}.

During evolution with two-local Trotter gates, the Vidal gauge is implicitly maintained throughout the time-evolution, however this is not the case when applying tensor network operators in general, where more than one bond are updated simultaneously.
To correctly perform simple update, the state must be re-gauged to re-obtain bond matrices on the composite bonds of size $D \eta$, which can be then truncated quasi-optimally by discarding the smallest elements of these, assuming the state is close enough to the Vidal gauge~\cite{kalis_fate_2012}.
In doing so, it is assumed the rank-1 environment formed by the bond matrices accurately approximates the environment surrounding the bond being truncated, however this is uncontrolled.

In this work, we use a new variant of the simple update truncation method that avoids the need to re-gauge the enlarged network of $d^2 D^4 \eta^4$-sized tensors, similar to the single-site version of the fast-full update method~\cite{phien_fast_2015, phien_infinite_2015} and the iterative truncation procedure described in Ref.~\cite{czarnik_projected_2015}: after application of the \ac{cpepo}, immediately truncate all but a single bond in the network using the $D\eta\times D$ isometries $U_{\alpha}$ and $V_{\alpha}$ obtained for each bond from the previous timestep. 
The remaining bond is then truncated using the standard simple-update procedure, defining a new set of isometries and a new set of bond matrices for that particular bond.
Using the following measure of convergence:
\begin{equation}\label{eq:conv-cond}
	\delta^{[i]}<\max_{\alpha \in \Lambda} \norm*{ \lambda_{\alpha}^{[i]} - \lambda_{\alpha}^{[i - 1]}}
	,\end{equation}
where $\lambda^{[i]}_{\alpha}$ is the diagonal weight matrix on bond $\alpha$ after $i$-iterations, this process is iterated until $\delta^{[i]} < \epsilon_{\mathrm{su}}$ where we have found $\epsilon_{\mathrm{su}} = \num{1d-8}$ to be a useful convergence threshold.
This avoids the need to re-gauge the network before each truncation and reduces the complexity of the simple-update QR-decomposition from $\mathcal{O}(d^4 D^6 \eta^6)$ to $\mathcal{O}(d^4 D^6 \eta^3)$. 
We also found it useful for algorithm stability to fix the gauge of the \ac{ipepo} at the end of each time step.
This is described in more detail in Appendix.~\ref{app:gauge}.

\subsection{Computing Observables}

Observables can be obtained by first computing reduced density matrices with respect to a finite region of the infinite lattice.
To do so, the scalar tensor network formed by contracting the physical indices of each tensor must be approximately contracted using a suitable boundary or tensor renormalization method.
In this paper, the \ac{vumps} method was used to find a boundary \ac{mps} approximation to the upper and lower portions of the lattice around any given a row in combination with the left and right fixed points of the transfer matrices formed from the \ac{mps} and a single row of the lattice.
Once these have been calculated, reduced density matrices of any size can be obtained by approximating the required partial trace of the infinite lattice with these boundary tensors, as shown in Appendix~\ref{app:vumps}.
\section{\label{sec:results}Numerical Results}
\subsection{Dissipative Long-Range Ising Model}
As an initial technical benchmark of the method, we simulated the real time dynamics of the dissipative long-range transverse-field Ising model:
\begin{equation}\label{eq:long-range-ising}
	H = \frac{J}{\mathcal{N}(\alpha)}\sum_{j\neq k\in\Lambda}
	\frac{
		\sz{j}
		\sz{k}
	}
	{\norm{\mathbf{r}_j - \mathbf{r}_k}^{\alpha}} + h \sum_{j\in \Lambda}\sx{j}
	,\end{equation}
with dissipation occurring at a rate $\gamma$ represented by the single onsite Lindblad operator $\hat{L}_j = \hat{\sigma}_j^z$.
The parameter $J$ defines the total strength of the interactions between lattice sites, and $h$ the strength of the transverse field.
The denominator $\mathcal{N}(\alpha)$ is the so-called Kac normalization~\cite{kac_van_1963} which for $\alpha > 2$ is finite with the analytical value $\mathcal{N}(\alpha) = 4\zeta(\alpha/2)\beta(\alpha/2)$ where $\zeta(x)$ and $\beta(x)$ are the Riemann zeta and Dirichlet beta functions respectively~\cite{actor_evaluation_1990,kazemi_driven-dissipative_2023}.
This is used to rescale the \emph{total} strength of the interactions on any given lattice site to be independent of $\alpha$ ensuring that any changes to the dynamics when changing $\alpha$ are caused by the change in the range of the interactions only.
When $\alpha \leq 2$, $\mathcal{N}(\alpha)$ diverges and model is no longer well defined in the thermodynamic limit, so we do not consider that case here.
As $\alpha \to \infty$, the short range model is recovered and $\mathcal{N}(\alpha \to \infty) = 4$ coinciding with the coordination number of the lattice.

\subsubsection{\label{subsec:exact}Exactly solvable model}

\begin{figure}[t]
	\centering
	\vspace{1em}
	\includegraphics[width=\columnwidth]{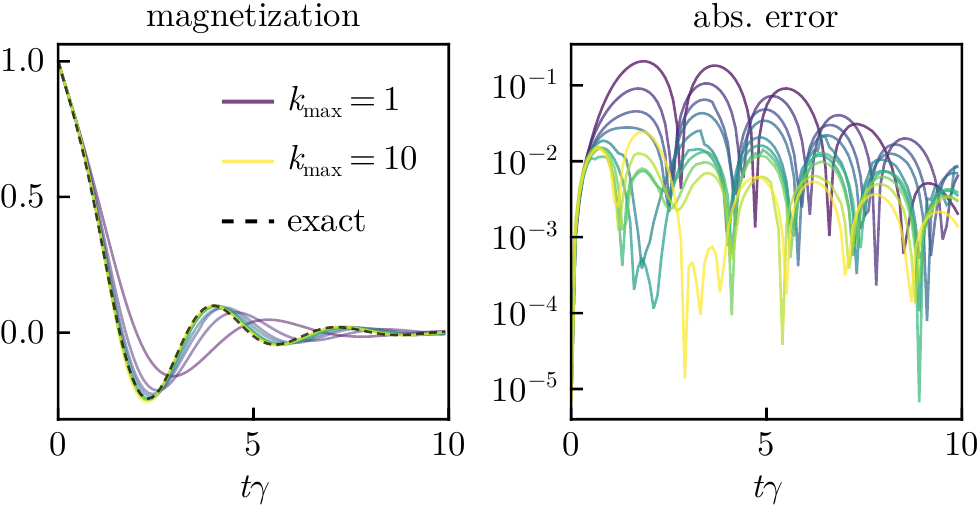}
	\caption{Local magnetization along the $x$-axis $m^x = \expval{\sigma_i^{x}}$ of the long-range Ising model as a function of time for varying values of $k_{\mathrm{max}}$ at bond dimension $D = 4$. The accuracy improves for increasing $k_{\mathrm{max}}$ as shown in the inset where the absolute different between the simulation and the exact solution (dashed) is plotted. The other hyper-parameters are $\max_{k} n_k = 4$, $g_{\mathrm{tol}} = 10^{-10}$ and $\Delta t = 0.01$.}
	\label{fig:kmax}
\end{figure}


In the absence of any transverse field, $h = 0$, the Lindblad equation~\eqref{eq:lindblad} with Hamiltonian~\eqref{eq:long-range-ising} has an exact solution.
We benchmark against this exact solution to assess how the \ac{fsa} approximation to the power-law interaction effects the accuracy of the simulation.
With $J /\gamma = 1.0$ and using an initial product state $\rho = \bigotimes \dyad{\uparrow_x}$, the model was simulated for a period of $t \gamma = 10 $.
The simulation was ran using a time step of $\Delta t = 0.01$ requiring a total of $1000$ time steps, with truncations performed using itrSU with
$\epsilon_{\mathrm{su}} = \num{1d-8} $ and at most $20$ iterations.
In constructing the approximation, $g_{\tol} = \num{1d-10}$.

\begin{figure}[t]
	\centering
	\includegraphics[width=\columnwidth]{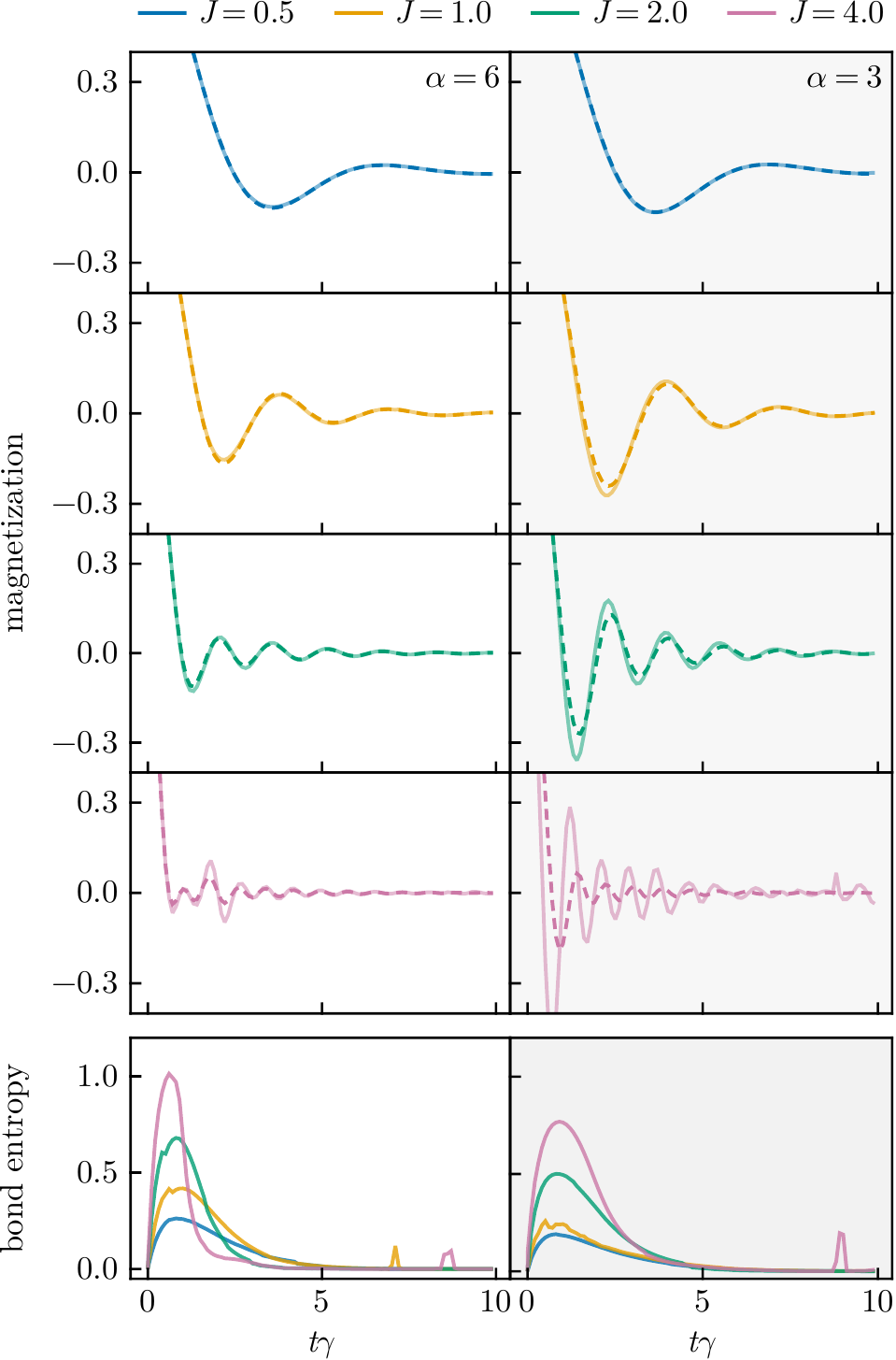}
	\caption{
    Comparison between the van der Waals case, $\alpha = 6$, (left, white background) and the dipolar case, $\alpha = 3$, (right, gray, background) of the long-range dissipative Ising model~\eqref{eq:long-range-ising} at varying interaction strength $J$. 
		 The bond entropy (two bottom panels) is larger  for increasing $J$, which is associated with worsening accuracy. The dissipation is set to $\gamma = 1$. 
        All other parameters are the same as in \figref{fig:kmax}. 
        }
	\label{fig:grid-j}
\end{figure}

First the effect of the cut-off $\kmax$ in the power-law approximation is investigated by considering comparing with the exact solutions.
As shown in \figref{fig:kmax}, the accuracy of the simulation, with respect to the exact solution, improves for increasing $\kmax$, with low values $\kmax < 5$ leading to visually inaccurate results.
For $\kmax = 10$, the results are similar to the exact solution with absolute error less than \num{1d-2} for the majority of the time evolution.
While, $\kmax = 10$ leads to the lowest absolute error during the late-time dynamics, defined here as $t\gamma \gtrsim 3.0$, before this, where the magnetization changes most rapidly, more accurate results can be achieved with smaller values of $\kmax$. 
This is likely a consequence larger $\kmax$ requiring more Trotter steps and thus more truncations which may not be found optimally via the uncontrolled simple update approximation.
This is supported by an increased entanglement during this period as evidenced by the peak in bond entropy plotted in \figref{fig:grid-j}.

We also investigated how the simulation performs when reducing the dissipation, equivalent to increasing the interaction strength.
It is expected that as the dissipation decreases, the entanglement will increase at a faster rate, compromising the accuracy of the finite bond-dimension tensor network simulation at early times~\cite{keever_stable_2021}.
To estimate this amount of entanglement in the system, we use the bond entropy,
\begin{equation}\label{eq:bond-entropy}
	S_{\alpha} = -\sum_{j = 1}^D [\lambda_{\alpha}]_{jj}\log[\lambda_{\alpha}]_{jj}
	,\end{equation}
which acts as a proxy to the true entanglement.
For acyclic tensor networks, the quantity $S_{\alpha}$ defined in~\eqref{eq:bond-entropy} reduces to the true entanglement entropy between the two regions of the network defined by cutting the bond $\alpha$.

Fixing $\kmax = 20$ and $\nmax = 4$, \figref{fig:grid-j} plots the magnetization for a range of interaction strengths $J /\gamma \in \{0.5,1.0,2.0,4.0\} $.
The bottom panels of \figref{fig:grid-j} show the peak in the bond entropy at a higher value for increasing $J/\gamma$, where the simulated dynamics tends to deviate the most from the exact solution.
Interestingly, the less-accurate $\alpha= 3$ simulations are associated with a shorter but broader peak in bond entropy suggesting that more prolonged periods of high bond entropy are more problematic than those with sharp but brief peaks.

\subsubsection{\label{subsec:beyond}Beyond exactly solvable regime}

\begin{figure*}
	\centering
	\includegraphics[width=\linewidth]{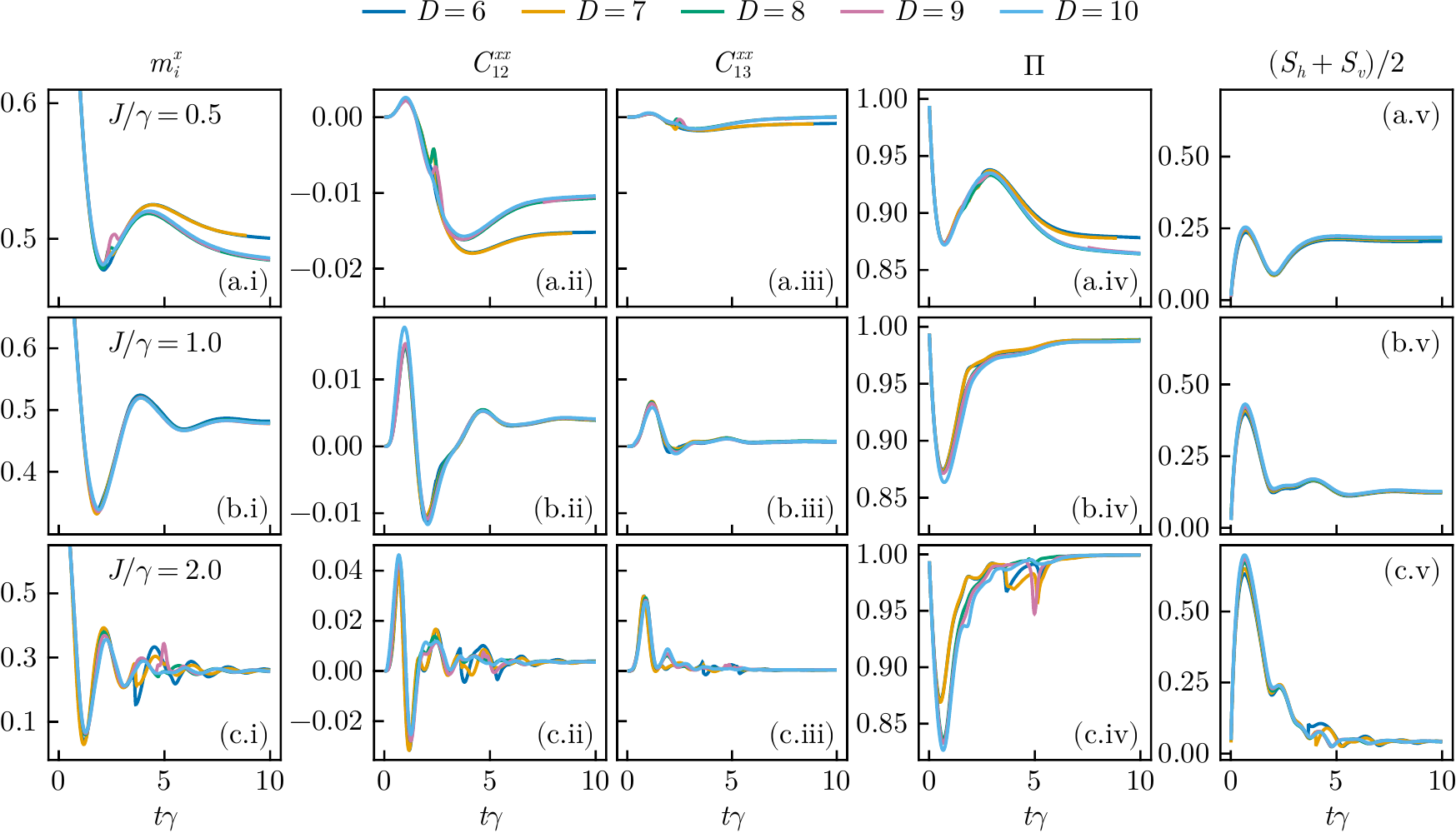}
	\caption{Plots of (i) magnetization $m_i^{x} = \la \sx{i} \ra$, correlators (ii) $C^{xx}_{12}$ and (iii) $C^{xx}_{13}$ where $C^{xx}_{ij} = \la \sx{i}\sx{j} \ra - \la \sx{i} \ra \la \sx{j} \ra$, (iv) purity, and (v) bond entropy of the long-range transverse field Ising model with $\alpha = 3.0$, $h / \gamma = 0.5$ and $J/\gamma$ equal to (a) $0.5$, (b) $1.0$ (c) $2.0$.
  A time step of $\Delta t= \num{1.25d-2}$ was used and bonds were truncated using parallel itrSU with a tolerance of $\num{1d-8}$. In constructing the approximation to the power law, $\kmax = 10$ Gaussians were used with $g_{\tol} = \num{1d-5}$ and $n_{\max} = 4$.} 
	\label{fig:ising-beyond}
\end{figure*}
To assess the performance of the method in a regime not yet amenable to an exact solution, the model~\eqref{eq:long-range-ising} is simulated with transverse magnetic field of strength $h/\gamma = 0.5$.
In this case, there is no known exact solution to the model~\eqref{eq:long-range-ising}.
Compared to the exactly solvable case, the solution appears to move toward a non-trivial steady-state with non-zero bond entropy, non-zero correlations and finite magnetization; the results are shown in Fig.~\ref{fig:ising-beyond}.
When $J/\gamma = 0.5$, the long time dynamics are effected by the bond dimension most significantly, with results $D \geq 8$ visually agreeing with each other, but visually distinct from the simulations with $D = 6$ and $D = 7$.
This can be seen in panels (a.i--a.iv) in Fig.~\ref{fig:ising-beyond}.
The time dynamics are smoothest for bond dimension $D = 10$ in all panels, however good convergence is found for lower bond dimension $D < 10$ for the case of $J/\gamma = 1.0$.
For the case $J / \gamma = 2.0$, the time dynamics are more difficult to converge with respect to the bond dimension, with some unstable regions in time for $D \in \{ 6,7 \}$, however simulations with $D = 10$ are smooth for all computed quantities. 
This demonstrates that the method is applicable to models with long-range interactions beyond the exactly solvable regime, provided care is taken to ensure convergence with bond dimension.



\subsection{\label{subsec:rydberg}Rydberg Hamiltonian}

\begin{figure*}[t]
	\centering
	\includegraphics{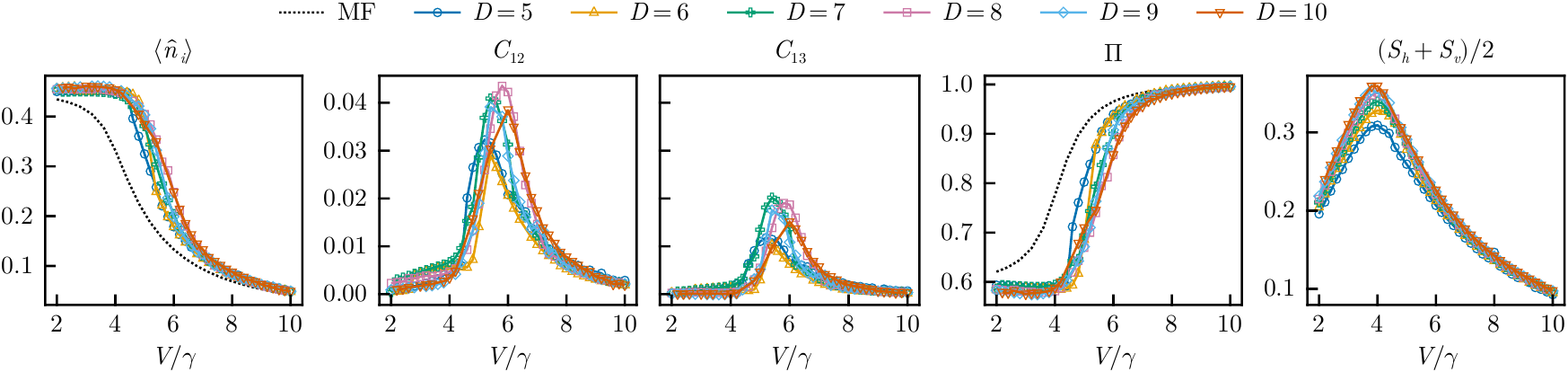}
	\caption{Steady state of the Rydberg Hamiltonian~\eqref{eq:rydberg} with $\Omega/\gamma = 2.0$ and $\Delta = -V/2\gamma$ at bond dimension $D \in 5..10$. The mean-field (dashed line) is plotted for observables where this is non-trivial. From left to right: the average occupation $\expval{\hat{n}_i }$, nearest-neighbor and next-nearest neighbor correlations $C_{jk}= \expval{\hat{n}_j \hat{n}_k}  - \expval{\hat{n}_j }\expval{\hat{n}_k} $, average purity $\Pi$, and the bond entropy $(S_h + S_v)/2$. }
	\label{fig:rydberg-steady-state}
\end{figure*}
\begin{figure*}[t]
	\centering
	\includegraphics{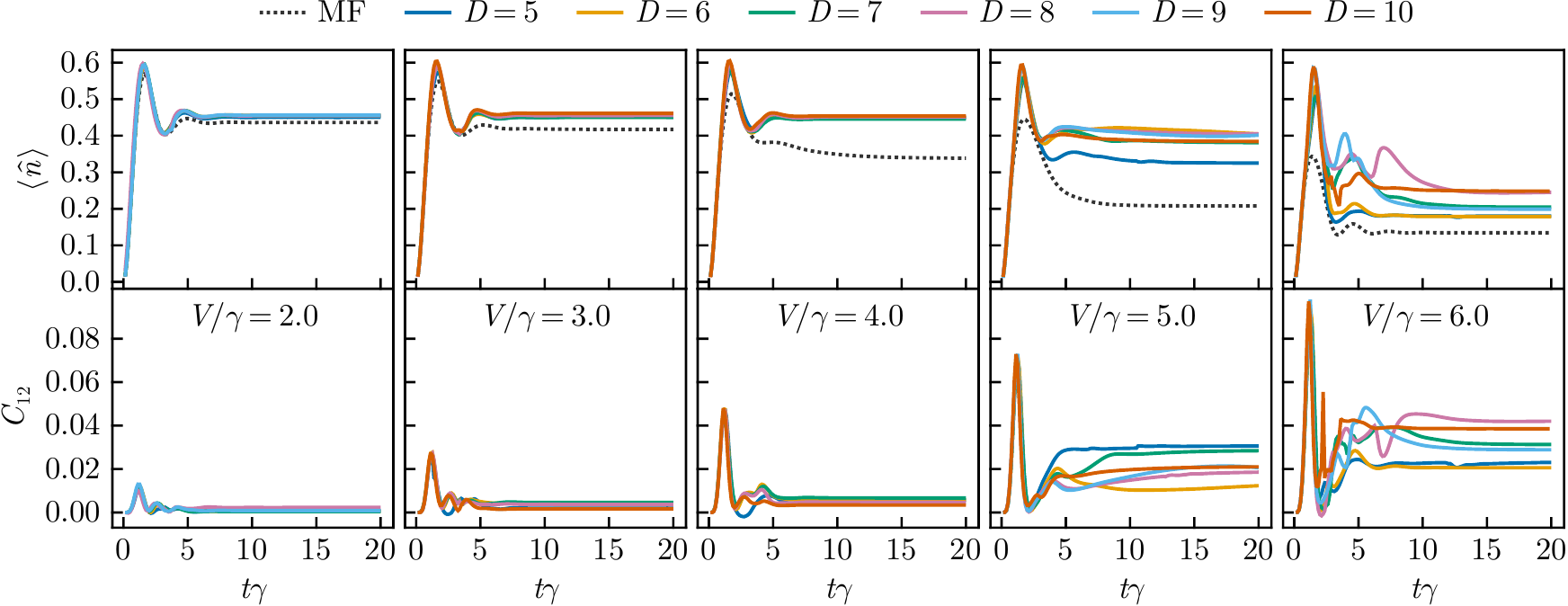}
	\caption{Time dynamics of the Rydberg Hamiltonian~\eqref{eq:rydberg} with $\Omega/\gamma = 2.0$ and $\Delta = -V/2\gamma$ at bond dimension $D \in 5 \isep 10$. The mean-field solution (dashed line) is plotted along side for observables where this is non-trivial.
	\label{fig:rydberg-dynamics-all}}
\end{figure*}

Finally, we consider a physically-motivated example.
The long-range Ising model can be realized experimentally using an array of atoms that can in either an excited Rydberg state $\ket{r}$ or a ground state $\ket{g}$.
When excited, each atom creates a strong short-range repulsive potential that is typically of van der Waals type $r^{-6}$.
However, for certain species of Rydberg atoms, the underlying $r^{-3}$ dipole-dipole interaction can be recovered via an effect known as F\"orster resonance~\cite{ravets_coherent_2014}, leading to a dipole-dipole blockading effect useful for quantum information processing~\cite{lukin_dipole_2001}.
Including a laser driving field and local dissipation from the excited state to ground state~\cite{kazemi_driven-dissipative_2023}, we consider the following model Hamiltonian:
\begin{equation}\label{eq:rydberg}
	H = \frac{V}{\mathcal{N}(\alpha)}\sum_{j\neq k \in \Lambda}\frac{\hat{n}_{j} \hat{n}_{k}}{\norm{\mathbf{r}_j - \mathbf{r}_k}^{\alpha}}
	- \frac{\Omega}{2}\sum_{j\in \Lambda}\sx{j}  + \Delta \sum_{j\in \Lambda} \hat{n}_j
\end{equation}
together with Lindblad operator $L_j = \sqrt{\gamma}\dyad{g_j}{r_j}$ evolving according to the Lindblad equation~\eqref{eq:lindblad}, where $\hat{n}_j = \dyad{r_j}$ and $\hat{\sigma}^x_j = \dyad{g_j}{r_j} + \dyad{r_j}{g_j}$.
The parameter $V/\mathcal{N}(\alpha)$ represents the Kac-normalized repulsion strength, $\Omega$ the Rabi frequency, $\Delta$ the laser de-tuning, and $\gamma$ the dissipation rate.
Variants of this model have been studied on a one-dimensional lattice with tree-tensor networks~\cite{sulz_numerical_2024}, however that particular ansatz is unable to capture area-law when applied to two-dimensional systems in general, and via a variational Monte Carlo applied to a matrix-product state ansatz~\cite{hryniuk_tensor-network-based_2024}.
In two-dimensions, the steady-state of the van der Waals case, $\alpha = 6$, has been studied using variational methods~\cite{kazemi_driven-dissipative_2023}, and in higher-dimensions using a semi-classical phase-space approach~\cite{huber_realistic_2022}.

\subsubsection{Steady state}

In \figref{fig:rydberg-steady-state} we plot the steady state at bond dimensions in the interval $D \in 5 \isep 10$.
A time step of $\Delta t= \num{1.25d-2}$ was used, and the simulation was ran for $2000$ iterations.
The parallel version of iterative simple update method was used to truncate the bonds after the application of each \ac{cpepo} operator.

We observe crossover from a phase with high average population $\expval{\hat{n}_j}$ to a phase with low average population as the strength of the interactions increase, characteristic of the dipole-dipole blockading affect.
This crossover occurs at $V/\gamma\approx 6.0$ where there is also an increase in the correlations $C_{jk} = \expval{\hat{n}_j \hat{n}_k} - \expval{\hat{n}_j} \expval{\hat{n}_k} $ between nearest-neighbor and next-nearest-neighbor sites ($C_{12}$ and $C_{13}$ respectively) in the lattice.
There is also associated with a transition from a mixed state with purity $\Pi \approx 0.6$ to a pure state with $\Pi \to 1.0$ as the dissipation rate $\gamma$ is deceased.
At large $V / \gamma $ the steady state becomes pure and one again well approximated by the mean field solution.

The sum of the bond entropy on each bond $(S_h +S_v)/2$ appears to peak at $V/\gamma \approx 4.0$ where the simulated result begins to deviate most significantly from the mean-field solution.
This is expected considering the limit of zero bond entropy resulting from a product state with all bond weights having only a single non-zero singular value.

\subsubsection{Dynamics}
In Figure~\ref{fig:rydberg-dynamics-all}, the real time dynamics at low-to-moderate interaction strength  $V/\gamma \in \{2.0, 3.0, 4.0, 5.0, 6.0\}$, is plotted.
Stable time-evolution is achieved when $V/\gamma \leq 5.0$ showing significant corrections to the mean field solution, also plotted. 
At $V\gamma = 6.0$, the time dynamics begin to deteriorate as the crossover point is traversed. 
Near this crossover point, the simulation deviates from the mean-field significantly and there is an associated increase in the correlations between neighboring lattice sites.

\begin{figure}[t]
	\centering
	\includegraphics[width=\columnwidth]{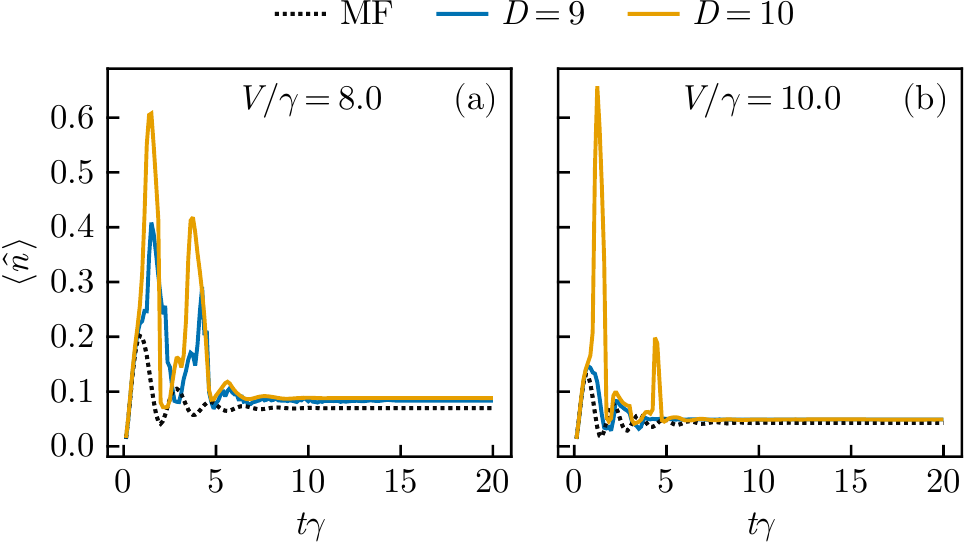}
	\caption{Dynamics of the Rydberg Hamiltonian~\eqref{eq:rydberg} in the weakly dissipative regime simulated using the \ac{cpepo} method at bond dimension $D = 9$ and $D = 10$. While dynamics are unstable at short times, convergence to a steady state at later times is not compromised.}
	\label{fig:rydberg-dynamics-high}
\end{figure}

At high interaction strength, $V\gamma \in \{8.0,10.0\}$, the rapid dynamics in the initial stages of the time evolution $t\gamma < 5.0$ leads to poor convergence of the trace environment for larger $V/\gamma$ leading to unreliable results during this time period.
This is shown in by the larger imaginary part of the average occupation, shown in Figure.~\ref{fig:rydberg-dynamics-high}.
However, we still find smooth evolution for at late time, and the initial unstable time evolution at $t < 5.0$ does not appear to compromise the simulation at later times; the simulation stabilizes and convergence to a steady state is still achieved.




\subsubsection{Comparison with short-range model}
A corresponding short-range (nearest-neighbor) model can be recovered in the limit $\alpha \to \infty$.
In this limit, we found that simulations with existing tensor network methods~\cite{kshetrimayum_simple_2017} fail to converge to a steady state and can be unstable in the worst case, with possibly spurious limit cycles emerging for certain simulation parameters.
The \ac{cpepo} method of this work is not expected to perform any better in this case, however it allows for simulation of the long-range regime beyond nearest-neighbor $2 <\alpha < \infty$.
This is in contrast to the short-range model, simulated using TEBD and standard simple update, where no steady state is achieved; by increasing the range of the interactions from short-range to dipolar, the time dynamics become more stable and convergence to steady states can be achieved for a much wider range of interaction strength $V/\gamma$.

\begin{figure}
	\centering
	\includegraphics[width=\columnwidth]{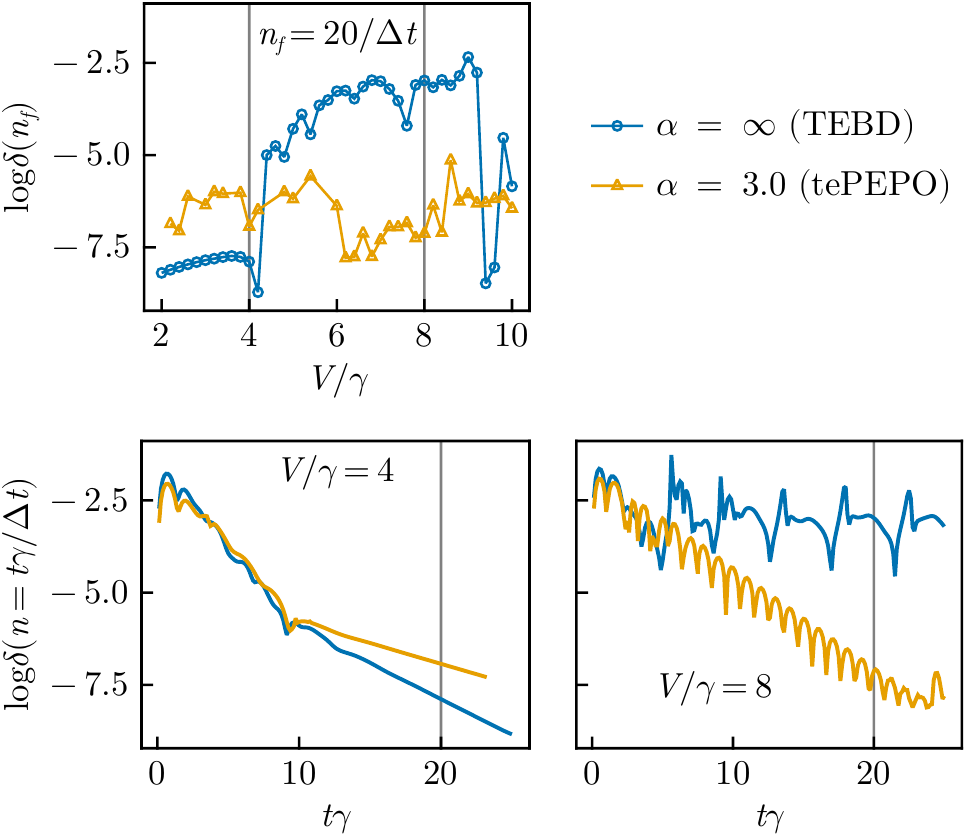}
	\caption{Steady state convergence at $D = 10$ using $\tau = \num{1.25d-2}$ for the \ac{cpepo} method compared to the short range model ($\alpha \to \infty$) simulated with TEBD.
		For the short range model, convergence cannot be achieved for much of the region $V/\gamma > 4$, whereas the simulation of $\alpha = 3.0$ using \ac{cpepo} are able to converge in this regime.
		For $V/\gamma \leq 4$, both cases converge. }
	\label{fig:rydberg-conv}
\end{figure}

To measure convergence in time, we use the quantity $\delta^{[n]}$ defined in~\eqref{eq:conv-cond} rescaled according to the time step $\delta^{[n]} \to \delta^{[n]} / \Delta t$, with $n$ referring to the $n$-th time step of the simulation such that $t = n \Delta t$.
As shown in Fig.~\ref{fig:rydberg-conv}, the short range model typical fails to convergence in time when $V/\gamma > 1.0$.
On the other hand, the long-range model ($\alpha = 3.0$) solved with the \ac{cpepo} method converges similarly for all values of $V/\gamma$ despite worse convergence in parameter regions where convergence can be found in the short range model.
This suggests the existence of long-range Lindblad equations that are amenable to simulations with tensor networks despite the corresponding short-range model failing to converge.

\section{Discussion}

We have proposed a construction of tensor network operators able to simulate the real and imaginary time dynamics of two-dimensional quantum lattice system, and have benchmarked the method against the extreme example of systems with power-law long-range interactions,  where we described how such interaction profiles can be approximated as sums of simpler terms.
We show that only a modest number of terms in this sum are required for visibly accurate results.
Our new simple-update based iterative truncation method is capable of efficient bond truncations when evolving using tensor network operators.

As with most high-dimensional tensor network methods, the core computational bottleneck remains the bond truncation step.
Despite it's efficiency, itrSU lacks the robustness of the regular simple update method, while still relying on a
good approximation to the Vidal gauge being achievable, which may not always be the case.
Furthermore, itrSU assumes the same uncontrolled approximation to the environment.
We found that the correlation length $\xi$ was on the order of a few lattice sites $\xi \gtrapprox 2$, suggesting that correlations extend beyond that which can be captured by bond weights in the simple update method.
Therefore, the \emph{environment} update, such as the fast full update~\cite{phien_fast_2015,phien_infinite_2015,czarnik_projected_2015}, will potentially improve the stability and accuracy of our method in these regions, where the correlation length becomes larger.
Another extension could be to truncate using the recently developed belief propagation method~\cite{tindall_gauging_2023}, where the usually uncontrolled simple update approximation can be systematically improved by including loop corrections~\cite{evenbly_loop_2025}.

The large class of Hamiltonians (and indeed Lindblad operators) that permit time evolution operators in terms of \ac{fsa} remains largely unexplored.
The extension to beyond on-site dissipation is straightforward, and would simply involve writing the three terms appearing in the vectorization of each Lindblad operator as \ac{fsa} rules, as shown in Appendix~\ref{app:lindblad-cons}.
Non-trivial dissipation has been associated with a wide range of quantum phenomena such as quantum time crystals~\cite{iemini_boundary_2018}, and quantum state preparation~\cite{wampler_absorbing_2024}, to name a few.
Finally, the examples presented in this paper are two-level many-body systems.
The extension to larger local Hilbert spaces is trivial and to fermions is also in principle straightforward.
In the former one has the advantage that the bond dimension of the operator is decoupled from dimension of the local Hilbert space. The path structure of the \ac{fsa} also allows encoding mediated cluster operators, such as those obtained from the Jordan-Wigner transformation, at no extra cost to the operator bond dimension.
In summary, this work represents a major advancement in simulating the dynamics of higher-dimensional quantum systems with tensor networks, and opens up new avenues in the study of novel many-body phenomena in both open and finite temperature quantum systems, as well as realistic quantum devices.

	{\it Acknowledgments.}
The authors thank Dawid Hynriuk for valuable discussions and feedback on the manuscript. 
This work was supported by the Engineering and Physical Sciences Research Council [grant number EP/S021582/1], and the European Union under Horizon Europe (project No. 101186579).
The authors acknowledge the use of the
UCL HPC facilities and associated support services in
the completion of this work.

\section*{DATA AVAILABILITY}
The data that support the findings of this article are openly available~\cite{dunham_2025_17656737}.

\bibliography{references}

\appendix
\section{\label{app:gauge}Gauge Fixing}

\begin{figure}
    \centering
    \includegraphics[width=\columnwidth]{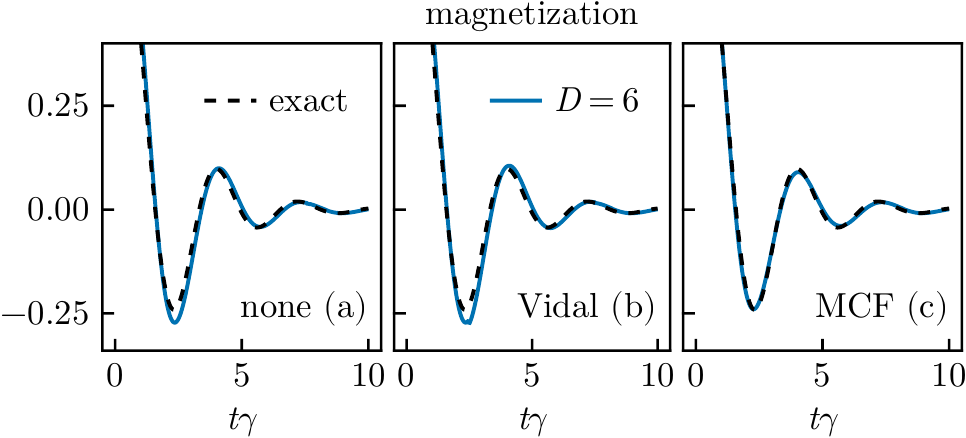}
    \caption{Effect of gauging fixing on the accuracy of the dissipative Ising model with $J = 1.0$. Plotted alongside is the exact solution (dashed).}
    \label{fig:gauge-fixing}
\end{figure}

The iterative stage in \ac{itrsu} implicitly moves the state to toward the Vidal gauge in a similar way to the standard simple update procedure over the course of many time-steps.
We found that in addition, explicitly gauging the tensor network to the Vidal gauge at each time step had little effect.
Interestingly however, we found more accurate results were obtained when fixing the gauge to the so-called \emph{minimal canonical form} described in Ref.~\cite{acuaviva_minimal_2023}, when compared to the exactly solvable model~\eqref{eq:long-range-ising}, as shown in \figref{fig:gauge-fixing}.

\section{\label{app:extra}Additional \ac{fsa} Examples}
\subsection{Symmetric rules for the $J_1$-$J_2$ Hamiltonian}
The \ac{fsa} rules that generate a given Hamiltonian are not unique.
For example, the $J_1$-$J_2$ Hamiltonian~\eqref{eq:ham-j1j2} can be represented using a \ac{cpepo} that is symmetric about both lattice axis at a cost of increasing the bond dimension by 1 on the vertical axis where the unconventional path is not necessary.
These rules are listed in \tabref{tab:fsm-j1j2-sym}.
\subsection{Toric code}
Hamiltonians with terms that are not strictly two-body can also be represented as \ac{fsa}. For example, the toric code Hamiltonian with qubits defined on the \emph{edges} of a lattice $\Lambda$ is given by
\begin{equation}\label{eq:toric}
H_{\mathrm{toric}} =  -J \sum_{v \in \Lambda}\left(\prod_{j\in v} \sx{j} \right)
- K\sum_{p \in \Lambda} \left( \prod_{j\in p}\sz{j} \right)
,\end{equation}
where $v \in \Lambda$ refers to the four qubits on the edges adjacent to the vertex $v$, and $p \in \Lambda$ refers to the four qubits on the edges defining the plaquette (face) $p$ of the lattice $\Lambda$. 
The Hamiltonian~\eqref{eq:toric} has an A/B sublattice checkerboard structure, with time-evolution operator generated by the rules give in \tabref{tab:fsm-toric}.

\begin{table}
	\renewcommand{\arraystretch}{1.5}
	\centering
	\begin{ruledtabular}
		\begin{tabular}{c  c  c  c}
			Rule                  & \makecell{Index                                           \\ $(e,s,w,n)$} & Value & Rule type \\
			\midrule
			$\fsmbox{0}{0}{0}{0}$ & $(0,0,0,0)$     & $\I$                         & identity \\
			$\fsmdot{0}{0}{0}{0}$ & $(0,0,0,0)$     & $-i\hat{D}$                    & D        \\
			\midrule
			$\fsmbox{0}{0}{2}{0}$ & $(0,0,2,0)$     & $\sqrt{-iJ_1} \hat{X}$       & C        \\
			$\fsmbox{0}{0}{0}{2}$ & $(0,0,0,2)$     & $\sqrt{-iJ_1} \hat{X}$       & B        \\
			\midrule
			$\fsmbox{1}{1}{0}{0}$ & $(1,1,0,0)$     & $J_2 J_1^{-1}\I$             & A        \\
			$\fsmbox{0}{1}{2}{0}$ & $(0,1,2,0)$     & $\frac{1}{2} J_2 J_1^{-1}\I$ & A        \\
			$\fsmbox{1}{0}{0}{2}$ & $(1,0,0,2)$     & $\frac{1}{2} J_2 J_1^{-1}\I$ & A        \\
			$\fsmbox{0}{0}{1}{0}$ & $(0,0,1,0)$     & $\sqrt{-iJ_1} \hat{X}$       & B        \\
			$\fsmbox{0}{0}{0}{1}$ & $(0,0,0,1)$     & $\sqrt{-iJ_1} \hat{X}$       & C        \\
			\midrule
			$\fsmbox{2}{0}{0}{0}$ & $(2,0,0,0)$     & $\sqrt{-iJ_1} \hat{X}$       & B        \\
			$\fsmbox{0}{2}{0}{0}$ & $(0,2,0,0)$     & $\sqrt{-iJ_1} \hat{X}$       & C        \\
		\end{tabular}
	\end{ruledtabular}
	\caption{
    Example of non-zero \ac{fsa} rules for the real time-evolution of the $J_1$-$J_2$ Ising-type Hamiltonian~\eqref{eq:ham-j1j2} such that the resulting \ac{cpepo} operator has symmetry between the lattice axes. The bond dimension operator is now $\eta_{h} =\eta_v =\eta = 3$ for all virtual bonds. We assume $J_1 > 0 $, however to treat $J_1 = 0$, remove the two rules in the second group and set $J_1 =1$.
        These rules should be used in combination with~\tabref{tab:prefactor} and Algorithm~\ref{alg:wii}.
		\label{tab:fsm-j1j2-sym}}
\end{table}

\begin{table}
	\renewcommand{\arraystretch}{1.5}
	\centering
	\begin{ruledtabular}
		\begin{tabular}{c  c  c  c}
			Rule                  & \makecell{Index                                           \\ $(e,s,w,n)$} & Value & Rule type \\
			\midrule
			$\fsmbox{0}{0}{0}{0}$ & $(0,0,0,0)$     & $\I$                         & identity \\
			\midrule
            \multicolumn{4}{l}{Sublattice A}\\
			$\fsmbox{1}{1}{0}{0}$ & $(1,1,0,0)$     & $\sqrt{-iJ} \hat{X}$       & C        \\
			$\fsmbox{0}{0}{2}{2}$ & $(0,0,2,2)$     & $\sqrt{-iJ} \hat{X}$       & B        \\
            $\fsmbox{2}{0}{0}{1}$ & $(2,0,0,1)$     & $\hat{Z}$       & A        \\
			$\fsmbox{0}{2}{1}{0}$ & $(0,2,1,0)$     & $\hat{Z}$       & A        \\
			\midrule
            \multicolumn{4}{l}{Sublattice B}\\
			$\fsmbox{1}{1}{0}{0}$ & $(1,1,0,0)$     & $\sqrt{-iK} \hat{Z}$       & C        \\
			$\fsmbox{0}{0}{2}{2}$ & $(0,0,2,2)$     & $\sqrt{-iK} \hat{Z}$       & B        \\
            $\fsmbox{2}{0}{0}{1}$ & $(2,0,0,1)$     & $\hat{X}$       & A        \\
			$\fsmbox{0}{2}{1}{0}$ & $(0,2,1,0)$     & $\hat{X}$       & A        \\
		\end{tabular}
	\end{ruledtabular}
	\caption{
        Example of non-zero \ac{fsa} rules for the real-time evolution of the toric code Hamiltonian~\eqref{eq:toric}. 
    The resulting \ac{cpepo} has an A/B sublattice checkerboard structure with bond dimension $\eta = 3$ on both lattice directions.
        These rules should be used in combination with~\tabref{tab:prefactor} and Algorithm~\ref{alg:wii}.
		\label{tab:fsm-toric}}
\end{table}

\section{\label{app:lindblad-cons}Explicit Construction of Vectorized Super-Operators}
Neglecting one-local terms for convenience, consider operators of the form
\begin{equation}
	H = \sum_{i,j\in \Lambda} \hat{C}_i \hat{B}_j 
	.\end{equation}
Under vectorization, the coherent part of the Lindblad master equation becomes:
\begin{equation}
	-i[H, \rho] \overset{\mathrm{vec}}{\to} \left[ \sum_{i,j\in \Lambda}
		\left(\hat{C}^{[+]}_j \hat{B}^{[+]}_k + \hat{C}^{[-]}_j \hat{B}^{[-]}_k \right) 
        \right] \vecp
	.\end{equation}
where we have defined the following terms on the Liouville space:
\begin{align}
	\hat{C}^{[+]}_j \hat{B}^{[+]}_k & = -i ( \I \otimes \hat{C})_j (\I \otimes
	\hat{B})_k,                                                                \\
	\hat{C}^{[-]}_j \hat{B}^{[-]}_k & = i ( \hat{C}^{\top} \otimes \I )_j (
	\hat{B}^{\top} \otimes \I)_k,
\end{align}
Each two-body term then contributes a value of $2$ to the bond dimension per term in the Hamiltonian.
Dissipation between different lattice sites can be treated by writing the terms in the dissipator a sum of distinct operators.
Consider the two-body Lindblad operator $\hat{L}_{i,j} = \sqrt{\gamma} \,\hat{\Gamma}_i \hat{\Gamma}_j$ acting on sites $i$ and $j$.
Under vectorization, this generates the following three two-body terms terms:
\begin{align}
	\hat{C}^{[1]}_j \hat{B}^{[2]}_k & = \gamma ({\hat{\Gamma}}^{*}\otimes\ \hat{\Gamma})_j (\hat{\Gamma}^{*}\otimes\ \hat{\Gamma})_k, \\
	\hat{C}^{[2]}_j \hat{B}^{[2]}_k & =-\frac{\gamma}{2}
	(\I \otimes \hat{\Gamma}^{\dagger}\hat{\Gamma})_j(\I \otimes \hat{\Gamma}^{\dagger}\hat{\Gamma})_k,                               \\
	\hat{C}^{[3]}_j \hat{B}^{[3]}_k & =-\frac{\gamma}{2}
	(\hat{\Gamma}^{\top}\hat{\Gamma}^{*} \otimes \I)_j(\hat{\Gamma}^{\top}\hat{\Gamma}^{*}\otimes\I)_k,
\end{align}
which can be represented using \ac{fsa} in the same way as done with Hamiltonian terms.
In general, each non-purely-local Lindblad operator thus contributes a value of $3$ to the total bond-dimension of the resulting tensor-network time-evolution operator.
The total generator of the time evolution can then be expressed as
\begin{multline}
	\mathcal{G} =\sum_{i,j\in\Lambda}\left[ \sum_k^{\phantom{3}} \bb{
	\hat{C}^{[k, +]}_i \hat{B}^{[k,+]}_j
	+\hat{C}^{[k, -]}_i \hat{B}^{[k,-]}_j
	}\right. \\ \left.
	+ \sum_{q}\sum_{\alpha=1}^{3}
	\hat{C}^{[q, \alpha]}_i \hat{B}^{[q,\alpha]}_j\right]
	.\end{multline}
\subsection{Thermal State Evolution}
Imaginary time evolution of a thermal state can be represented in a similar way to Lindblad real-time evolution.
A finite temperature state $\rho(\beta)$ can be expressed in terms of the Hamiltonian $H$ as
\begin{equation}\label{eq:thermal}
	\rho(\beta +\delta \beta) =
	e^{-\frac{\delta \beta}{2} H}
	\rho(\beta )
	e^{-\frac{\delta \beta}{2} H}
	,\end{equation}
where $\beta = 1/k_{\mathrm{B}} T$ is the thermodynamic beta in terms of temperature $T$ and Boltzmann constant $k_{\mathrm{B}}$.
The vectorized from of the generator of the evolution~\eqref{eq:thermal} is then given by 
\begin{equation}
	\mathcal{G}_{\mathrm{th}} = -\frac{1}{2}\left(\I \otimes H  + H^{\top} \otimes \I \right)
	\label{eq:}
	.\end{equation}
Therefore, by defining:
\begin{align}
	\hat{C}^{[\mathrm{k}]}_j \hat{B}^{[\mathrm{k}]}_k & = -\frac{1}{2}( \I \otimes \hat{C})_j (\I \otimes
	\hat{B})_k,                                                                                           \\
	\hat{C}^{[\mathrm{b}]}_j \hat{B}^{[\mathrm{b}]}_k & = -\frac{1}{2} ( \hat{C}^{\top} \otimes \I )_j (
	\hat{B}^{\top} \otimes \I)_k,
\end{align}
the generator of the time-evolution can be expressed as
\begin{equation}
	\mathcal{G_{\mathrm{th}}} =\sum_{i,j\in\Lambda}\left[ \sum_k^{\phantom{3}} \bb{
	\hat{C}^{[k, \mathrm{k}]}_i \hat{B}^{[k,\mathrm{k}]}_j
	+\hat{C}^{[k, \mathrm{b}]}_i \hat{B}^{[k,\mathrm{b}]}_j
	}\right]
	,\end{equation}
with an additional arbitrary number of local terms that do not contribute to the bond dimension.
\section{\label{app:su}Iterative Simple Update}
Assuming the \ac{cpepo} time-evolution super operator has been applied to each \ac{ipepo} tensor in the lattice, then the the \ac{itrsu} method is summarized as follows:
Assume each bond $\alpha = (u,v)$ was truncated in the previous time-step by isometries $U_{\alpha}$ and $V_{\alpha}$ where $U_{\alpha}$ truncates the tensors at $u$ and $V_{\alpha}$ truncates the tensors at $v$.
Then each tensor at $v$ as an isometry $V^{[0]}_{u,\beta}$ associated to each bond $\beta$ incident to $u$ that was used to reduce the dimension of associated index of the tensors $A_v$ in the previous timestep.
Then the $n$-th iteration of \ac{itrsu} proceeds as follows:
\begin{enumerate}
	\item For bond $\alpha = (u,v)$, form the tensors $P_A$ and $P_B$ by respectively applying isometries $V^{[n - 1]}_{u, \beta}$ on $A_u$ and $ V^{[n - 1]}_{v, \gamma}$ on $A_v$ for bonds $\beta,\gamma \neq \alpha$, i.e. truncate the bonds of $A_u$ and $A_v$ excluding bond $\alpha$.
	\item Perform \emph{parallel simple update}, i.e.
	      \begin{enumerate}
		      \item absorb the square root of the weights $\lambda^{[n-1]}$ on bonds $\beta, \gamma \neq \alpha$ intro the appropriate tensors.
		      \item decompose tensors $P_A = Q_A M_A$ and $M_B Q_B = P_B$ using the QR decomposition, where the indices are grouped such that the only bonds in domain (codomain) of $P_A$ ($P_B$) are $\alpha$.
		      \item perform a truncated singular value decomposition on the product $M_A M_B$ to obtain isometries and new bond weight:
		            \begin{equation}
			            M_A M_B \overset{\mathrm{SVD}_D}{\rightarrow} U \lambda_\alpha^{[n]} V^{\dagger}
		            \end{equation}
		      \item construct the isometries:
		            \begin{align}
			            \tilde{U} & = M_A^{+} U \sqrt{\lambda_{\alpha}^{[n]}},
			            \\ \qq{and}\tilde{V} &= \sqrt{\lambda_\alpha^{[n]}}V M_B^{+},
		            \end{align}
		            where $M^{+}$ is the pseudo-inverse of $M$~\cite{moore_reciprocal_1920, bjerhammar_rectangular_1951, penrose_generalized_1955}.
	      \end{enumerate}
	\item Set new isometries $V_{u, \alpha}^{[n]} = \tilde{U}$ and $V_{v, \alpha}^{[n]} = \tilde{V}^{\dagger}$ and bond weight $\lambda^{[n]}_{\alpha}$.
    \item Repeat steps 1--3 for each bond in the lattice.
\end{enumerate}
This process is repeated until singular values $\lambda^{[n]}_{\alpha}$ converge to desired threshold.
Finally, once converged, truncated tensors $\tilde{A}_v$ are formed by truncating each bond of each tensor with $V^{[n]}_{v,\alpha}$.

The \ac{itrsu} method can be performed in parallel, where each site in the lattice is update independently at each iteration, or in sequence, where the isometries obtained from truncating a bond are used in subsequent truncations during the same iterative cycle.
The former is useful for maintaining lattice symmetries assuming a symmetric time-evolution operator.
We find consistent convergence is found using parallel \ac{itrsu} with a tolerance on the order of $\sim \num{1d-8}$.

\section{\label{app:vumps}VUMPS}

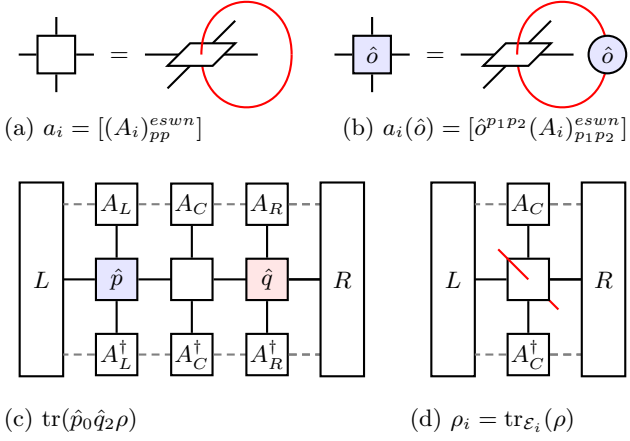
\begin{figure}[t]
	\centering
	\begin{equation*}
		\vcenter{\hbox{
				\begin{tikzpicture}[remember picture]
					\node[draw, thick, minimum size=0.5cm](A)at (0,0){};
					\draw[thick, ] (A) -- ++ (0,-0.50);
					\draw[thick, ] (A) -- ++ (-0.50,0);
					\draw[thick, ] (A) -- ++ (+0.50,0);
					\draw[thick, ] (A) -- ++ (0,+0.50);
					\coordinate (figA) at (current bounding box.south west) ;
				\end{tikzpicture}
			}}
		=
		\vcenter{\hbox{
				\begin{tikzpicture}
					\coordinate[inner sep=0pt] (A) at (0,0);
					\coordinate[inner sep = 0pt] (B) at (1.2,0);
					\draw[thick, red] (B.270) to [out = 270, in = 270, looseness=2] (A.270);
					\begin{scope}[scale=1.5, x={(1cm,0cm)},y={(0.5cm,0.5cm)},z={(0cm,1cm)}]
						\draw[thick,] (0,-0.20,0) -- ++ (0,-0.40,0);
						\draw[thick,] (-0.20,0,0) -- ++ (-0.30,0,0);
						\draw[thick,] (+0.20,0,0) -- ++ (+0.30,0,0);
						\draw[thick,] (0,+0.20,0) -- ++ (0,+0.40,0);
						\draw[thick, fill=white] (-0.20, -0.20, 0) -- ++ (0.4,0,0) -- ++ (0,0.4,0) -- ++(-0.4, 0,0) -- cycle;
					\end{scope}
					\draw[thick, red, ] (A.90) to [out = 90, in = 90, looseness=2] (B.90);
				\end{tikzpicture}
			}}
		\quad
		\vcenter{\hbox{
				\begin{tikzpicture}[remember picture]
					\node[draw, thick, minimum size=0.5cm, fill=blue!10](A) at (0,0){$\hat{o}$};
					\draw[thick, ] (A) -- ++ (0,-0.50);
					\draw[thick, ] (A) -- ++ (-0.50,0);
					\draw[thick, ] (A) -- ++ (+0.50,0);
					\draw[thick, ] (A) -- ++ (0,+0.50);
					\coordinate (figB) at (current bounding box.south west |- figA);
				\end{tikzpicture}
			}}
		=
		\vcenter{\hbox{
				\begin{tikzpicture}
					\coordinate[inner sep=0pt] (A) at (0,0);
					\coordinate[inner sep = 0pt] (B) at (1.2,0);
					\draw[thick, red, ] (A.270) to [out = 270, in = 270, looseness=2] (B.270);
					\begin{scope}[scale=1.5, x={(1cm,0cm)},y={(0.5cm,0.5cm)},z={(0cm,1cm)}]
						\draw[thick, ] (0,-0.20,0) -- ++ (0,-0.40,0);
						\draw[thick, ] (-0.20,0,0) -- ++ (-0.30,0,0);
						\draw[thick, ] (+0.20,0,0) -- ++ (+0.30,0,0);
						\draw[thick, ] (0,+0.20,0) -- ++ (0,+0.40,0);
						\draw[thick, fill=white] (-0.20, -0.20, 0) -- ++ (0.4,0,0) -- ++ (0,0.4,0) -- ++(-0.4, 0,0) -- cycle;
					\end{scope}
					\draw[thick, red, ] (A.90) to [out = 90, in = 90, looseness=2] (B.90);
					\node[draw ,circle, fill=blue!10, thick, minimum size=0.5cm, label=center:{$\hat{o}$}] (p) at (B) {};
				\end{tikzpicture}
			}}
		\label{eq:bulk-tensor-op}
	\end{equation*}
	\break
	\begin{equation*}
		\vcenter{\hbox{
				\begin{tikzpicture}[
						scale=1.0,tensor/.style ={draw, thick, fill=white, minimum size=0.55cm}
					]
					\node[tensor, opacity=0](UL) at (0,1){};
					\node[tensor, opacity=0](ML) at (0,0){};
					\node[tensor, opacity=0](DL) at (0,-1){};
					\node[tensor, opacity=0](UR) at (4,1){};
					\node[tensor, opacity=0](MR) at (4,0){};
					\node[tensor, opacity=0](DR) at (4,-1){};
					\node[draw, thick, fill=white, fit = (UL) (DL), inner sep = 0, label=center:{$L$}]{};
					\foreach \lab [count=\x] in {A_L, A_C, A_R} {
					\node[tensor, label=center:{$\lab$}](A\x) at (\x,1) {};
					\node[tensor, label=center:{$\lab^{\dagger}$}](B\x)at (\x,-1) {};
					\node[tensor, opacity=0] (\x) at (\x,0) {};
					\draw[thick,] (A\x) to (\x);
					\draw[thick,] (B\x) to (\x);
					\draw[thick,] (\x) -- ++ (0.75,0);
					\draw[thick,gray,densely dashed] (A\x) -- ++ (0.75,0);
					\draw[thick,gray,densely dashed] (B\x) -- ++ (0.75,0);
					}
					\node[tensor, fill=blue!10] at (1) {$\hat{p}$};
					\node[tensor] at (2) {};
					\node[tensor, fill=red!10] at (3) {$\hat{q}$};
					\draw[thick,] (1) -- (ML);
					\draw[thick,gray,densely dashed] (A1) -- (UL);
					\draw[thick,gray,densely dashed] (B1) -- (DL);
					\draw[thick,] (3) -- (MR);
					\draw[thick,gray, densely dashed] (A3) -- (UR);
					\draw[thick,gray, densely dashed] (B3) -- (DR);
					\node[draw, thick, fill=white, fit = (UR) (DR), inner sep = 0, label=center:{$R$}]{};
				\end{tikzpicture}
			}}
		\quad\quad
		\vcenter{\hbox{
				\begin{tikzpicture}[
						scale=1.0,tensor/.style ={draw, thick, fill=white, minimum size=0.55cm}
					]
					\node[tensor, opacity=0](UL) at (0,1){};
					\node[tensor, opacity=0](ML) at (0,0){};
					\node[tensor, opacity=0](DL) at (0,-1){};
					\node[tensor, opacity=0](UR) at (2,1){};
					\node[tensor, opacity=0](MR) at (2,0){};
					\node[tensor, opacity=0](DR) at (2,-1){};
					\node[draw, thick, fill=white, fit = (UL) (DL), inner sep = 0, label=center:{$L$}]{};
					\draw[thick, red] (1.center) --++ (0.4,-0.4);
					\foreach \x in {1} {
					\node[tensor, label=center:{$A_C$}](A\x) at (\x,1) {};
					\node[tensor, label=center:{$A^{\dagger}_C$}](B\x)at (\x,-1) {};
					\node[tensor, opacity=0] (\x) at (\x,0) {};
					\draw[thick,] (A\x) to (\x);
					\draw[thick,] (B\x) to (\x);
					\draw[thick,] (\x) -- ++ (0.75,0);
					\draw[thick,gray,densely dashed] (A\x) -- ++ (0.75,0);
					\draw[thick,gray,densely dashed] (B\x) -- ++ (0.75,0);
					}
					\node[tensor] at (1) {};
					\draw[thick, red] (1.center) --++ (-0.4,0.4);
					\draw[thick,] (1) -- (ML);
					\draw[thick,gray,densely dashed] (A1) -- (UL);
					\draw[thick,gray,densely dashed] (B1) -- (DL);
					\draw[thick,] (1) -- (MR);
					\draw[thick,gray, densely dashed] (A1) -- (UR);
					\draw[thick,gray, densely dashed] (B1) -- (DR);
					\node[draw, thick, fill=white, fit = (UR) (DR), inner sep = 0, label=center:{$R$}]{};
				\end{tikzpicture}
			}}
	\end{equation*}
	\begin{tikzpicture}[remember picture, overlay]
		\node[anchor=north west, yshift=-0.2cm, xshift=-0.3cm] (labA) at (figA) {(a) $a_i = [(A_i)_{pp}^{eswn}]$  };
		\node[anchor=north west, yshift=-0.2cm] (labB) at (figB) {(b) $a_i({\hat{o}}) = [\hat{o}^{p_1p_2} (A_i)_{p_1 p_2}^{eswn}]$  };
	\end{tikzpicture}
	\begin{tikzpicture}[remember picture, overlay]
		\node[anchor = west, yshift=-0.4cm] at (labA.west |- 0,0) {\strut (c) $\tr(\hat{p}_0\hat{q}_2\rho) $};
		\node[anchor = west, yshift=-0.4cm, xshift=0.9cm] at (labB.west |- 0,0) {\strut (d) $\rho_i = \tr_{\mathcal{E}_i}(\rho)$};
	\end{tikzpicture}
	\bigbreak
	\caption{(c) Example of how to compute a next-nearest neighbor correlation function using boundary MPS $\ket{\Psi(\{A_C,A_L,A_R\})}$ and transfer matrix fixed points $L$ and $R$ obtained from applying the \ac{vumps} procedure to the network formed from the rank-4 tensors $\{a\}$ defined in (a). In (d), the construction of a reduced density matrix is shown.
		The partial trace of the infinite lattice is approximated by an environment with finite bond-dimension $\chi$ shown by the dashed lines.
        \label{fig:vumps}
	}
\end{figure}
To compute $\tr(\rho)$, one must contract the entire infinite tensor network, including physical indices.
As this is not possible exactly, one uses an appropriate renormalization group or boundary algorithm to approximate this contraction.
In the context of computing observables, boundary methods are more suitable as it is more straightforward to embed non-translationally invariant tensors in the partially contracted tensor network. 
This is useful to, for example, construct reduced density matrices by embedding a intact \ac{ipepo} tensor $A_i$ who's physical indices are to remain \emph{un-traced} into a trace environment, $\mathcal{E}_i$ shown in \figref{fig:unit-cell}b, represented by the boundary tensors.

The method used in this is the paper is \ac{vumps} algorithm~\cite{zauner-stauber_variational_2018}, adapted to the generic task of computing contractions of infinite two-dimensional tensor networks~\cite{fishman_faster_2018,nietner_efficient_2020}.
The output of \ac{vumps} is a boundary \ac{mps} $\ket{\Psi(\{A_C,A_L,A_R\})}$ with bond-dimension $\chi$ that approximates the contraction of the upper and lower infinite portions of the lattice, and left and right fixed points $L$ and $R$ of the transfer matrices:
\begin{equation}
T_L
\coloneq
\vcenter{\hbox{
				\begin{tikzpicture}[
						scale=1.0,tensor/.style ={draw, thick, fill=white, minimum size=0.55cm},
                        font=\footnotesize,
					]
                \node[tensor, label=center:{$A_L$}](A0) at (0,1) {};
                \node[tensor, label=center:{$A_L^{\dagger}$}](B0)at (0,-1) {};
                \node[tensor] (0) at (0,0) {};
                \draw[thick,] (A0) to (0);
                \draw[thick,] (B0) to (0);
                \draw[thick,] (0) -- ++ (0.75,0) node[black, right] {$D$};
                \draw[thick,] (0) -- ++ (-0.75,0);
                \draw[thick,gray,densely dashed] (A0) -- ++ (0.75,0) node[black, right] {$\chi$};
                \draw[thick,gray,densely dashed] (B0) -- ++ (0.75,0) node[black, right] {$\chi$};
                \draw[thick,gray,densely dashed] (A0) -- ++ (-0.75,0);
                \draw[thick,gray,densely dashed] (B0) -- ++ (-0.75,0);
                \end{tikzpicture}}},\,\text{and}\quad
                T_R\coloneq
\vcenter{\hbox{
				\begin{tikzpicture}[
						scale=1.0,tensor/.style ={draw, thick, fill=white, minimum size=0.55cm},
                        font=\footnotesize,
					]
                \node[tensor, label=center:{$A_R$}](A0) at (0,1) {};
                \node[tensor, label=center:{$A_R^{\dagger}$}](B0)at (0,-1) {};
                \node[tensor] (0) at (0,0) {};
                \draw[thick,] (A0) to (0);
                \draw[thick,] (B0) to (0);
                                \draw[thick,] (0) -- ++ (0.75,0) node[black, right] {$D$};
                \draw[thick,] (0) -- ++ (-0.75,0);
              \draw[thick,gray,densely dashed] (A0) -- ++ (0.75,0) node[black, right] {$\chi$};
                \draw[thick,gray,densely dashed] (B0) -- ++ (0.75,0) node[black, right] {$\chi$};
                \draw[thick,gray,densely dashed] (A0) -- ++ (-0.75,0);
                \draw[thick,gray,densely dashed] (B0) -- ++ (-0.75,0);
                \end{tikzpicture}}}
,\end{equation}
i.e. $L \,T_L = L$ and $ T_R\, R = R$.

In~\figref{fig:vumps} we show how to compute the $n$-point correlation function $\expval{\hat{p}_0 \hat{q}_2} $, and a single site reduced density matrix using the \ac{vumps} boundary tensors.
Correlation functions between sites diagonal to each other are not compatible with \ac{vumps}. In that case one can make use of an alternative method such as the corner transfer matrix renormalization group~\cite{corboz_stripes_2011,fishman_faster_2018}.
\end{document}